\documentclass[a4paper,aps,prd,10pt,showpacs,twocolumn,nobibnotes,nofootinbib]{revtex4-1}
\usepackage{graphicx,amsmath,amsfonts,amssymb,revsymb,dcolumn,epsfig,bm,xspace}
\usepackage{hyperref}
\usepackage[latin1]{inputenc}
\usepackage{color}
\usepackage{bm}
\usepackage{mathtools}

\usepackage{mathrsfs}    
\usepackage{epsfig}      
\usepackage{accents}     
\usepackage{cancel}      
\usepackage{indentfirst} 
\usepackage[hang,splitrule]{footmisc} 
\usepackage{slashed} 
\usepackage[english]{babel} 
\usepackage{chngcntr} 
\usepackage{graphicx} 
\usepackage{tikz}
\usetikzlibrary{patterns,arrows,decorations.pathreplacing}
\usepackage{url}

\def\b{\beta}

\def\f{\phi}

\def\h{\eta}

\def\l{\lambda}

\def\p{\pi}

\def\x{\xi}

\newcommand{\ti}\tilde
\newcommand{\wt}\widetilde
\newcommand{\wh}\widehat
\newcommand{\bv}\breve
\newcommand{\dg}\dagger

\newcommand{\hs}{\hspace}

\newcommand{\Pl}{\mathrm{p}}
\newcommand*\dd{\mathop{}\!\mathrm{d}}
\newcommand{\Wrons}[2]{\mathcal{W}\left(#1,#2\right)}

\begin{document}

\title{Particle Creation in Bouncing Cosmologies}

\author{Diogo~C.~F.~Celani} \email{celani@cbpf.br}
\author{Nelson~Pinto-Neto} \email{nelson.pinto@pq.cnpq.br}
\author{Sandro~D.~P.~Vitenti} \email{vitenti@cbpf.br}

\affiliation{Centro Brasileiro de Pesquisas F\'{\i}sicas,
Rua Dr.\ Xavier Sigaud 150 \\
22290-180, Rio de Janeiro -- RJ, Brasil}

\date{\today}

\begin{abstract}
We investigate scalar particle creation in a set of bouncing models where
the bounce occurs due to quantum cosmological effects described by the
Wheeler-DeWitt equation. The scalar field can be either conformally or
minimally coupled to gravity, and it can be massive or massless, without
self interaction. The analysis is made for models containing a single
radiation fluid, and for the more realistic case of models containing the
usual observed radiation and dust fluids, which can fit most of the
observed features of our Universe, including an almost scale invariant
power spectrum of scalar cosmological perturbations. In the conformal
coupling case, the particle production is negligible. In the minimal
coupling case, for massive particles, the results point to the same
physical conclusion within observational constraints: particle production
is most important at the bounce energy scale, and it is not sensitive
neither to its mass nor whether there is dust in the background model.
The only caveat is the case where the particle mass is larger than the
bounce energy scale. On the other hand, the energy density of produced
massive particles depend on their masses and the energy scale of the
bounce. For very large masses and deep bounces, this energy density may
overcome that of the background. In the case of massless particles, the
energy density of produced particles can become comparable to the
background energy density only for bounces occurring at energy scales
comparable to the Planck scale or above, which lies beyond the scope of
this paper: we expect that the simple Wheeler-DeWitt approach we are
using should be valid only at scales some few orders of magnitude below
the Planck energy. Nevertheless, in the case in which dust is present,
there is an infrared divergence, which becomes important only for scales
much larger than today's Hubble radius.
\end{abstract}

\pacs{04.62.+v, 98.80.-k, 98.80.Jk}

\maketitle

\section{Introduction}

In cosmological bouncing scenarios, the initial singularity present in
the standard cosmological model is absent: the Universe contracts from a
very large size, bounces when it becomes sufficiently small, and expand
afterwards. The bounce can occur once or many times, depending on the
model. This fact (the absence of singularities) is, {\it{per se}},
already very important, and a sufficient reason to investigate these
models more deeply. Furthermore, other puzzles of the standard
cosmological model are absent in these scenarios (as the horizon and
flatness problems), and they can also supply a mechanism to generate
primordial cosmological perturbations from quantum vacuum fluctuations,
with almost scale invariant spectrum~\cite{Peter2007, Peter2008}, as in
inflationary models. Hence, they can also be viewed as alternatives to
inflation, although they are not necessarily contradictory to it.

There are nowadays many mechanisms to generate the bounce, and there are
also many open questions and issues to be investigated concerning these
models, for reviews see Refs.~\cite{Novello2008,Battefeld2015}. One of
them concerns the phenomenon of quantum particle creation around the
bounce: is it large enough to modify the background and/or induce some
sort of reheating, making the model asymmetric around the bounce, or it
is always negligible?

The aim of this paper is to investigate particle creation in the set of
bouncing models that have been investigated by some of us in the last
decades. In these models, the bounce occurs due to quantum cosmological
effects when the curvature of space-time becomes very large: the related
Wheeler-DeWitt equation is solved, and interpreted using the de
Brogli-Bohm quantum theory (the usual Copenhagen point of view cannot be
used in quantum cosmology, see Ref.~\cite{Pinto-Neto2013} for a review on
this subject). The trajectories describing the scale factor evolution are
calculated and they are usually non-singular, presenting a bounce due to
quantum effects at small scales, and turning to a classical standard
evolution when the scale factor becomes sufficiently large
\cite{AcaciodeBarros1998,Peter2007,Alvarenga2002,Pinto-Neto2013}. These
models usually contain one single hydrodynamical fluid, or two fluids
\cite{Pinto-Neto2005}: the usual observed radiation and dust contents
which are present in our Universe. We expect that the simple
Wheeler-DeWitt approach we are using should be valid only at scales some
few orders of magnitude below the Planck energy, being a limit of some
more involved theory of quantum gravity suitable for energy scales close
or above the Planck scale (see Refs.~\cite{Ashtekar2013,Ashtekar2014} as
examples of approaches in this direction).

In the models we will analyze, the Belinsky-Khalatnikov-Lifshitz (BKL)
instability \cite{Belinskii1970} is not addressed. Of course it would be
better to work out models without such instability. However, to insert an
extra ekpyrotic scalar field \cite{Khoury2001,
Buchbinder2007,Lehners2007} with suitable but ad-hoc potential seems
excessive to solve this problem, turning the advantage of this type of
models quite subjective. Note that any cosmological model, either
inflationary, bouncing, or any other approach to primordial cosmology,
has a much more serious problem to deal with: the large degree of initial
homogeneity necessary to turn all these models compatible with
observations (or at least to turn them computationally feasible). This is
infinitely more serious than the BKL problem, as long as one has to turn
identical the infinite many possible functions of time per space point
characterizing a general inhomogeneous gravitational field in order to
obtain a homogeneous geometry. If some unknown physical mechanism or
theory of initial conditions is capable to justify such extreme fine
tuned state, then it would not be a big surprise that it could also make
identical the three remaining time functions characterizing the three
directions of space. Once one has initially assumed a homogeneous and
isotropic universe, one can show for the models we are considering that
the shear perturbation will never overcome the background degrees of
freedom, even growing as fast as $a^{-6}$ in the contracting phase. This
is because the shear perturbation in such models is multiplied by a very
small number, a factor $(L_{\Pl} / (a_0R_H))^2$, where $L_{\Pl}$ is the
Planck length and $a_0R_H$ is the Hubble radius today, and if the bounce
is not very deep, the shear will always remain sufficiently small (see
Ref.~\cite{Vitenti2012, Vitenti2014} for details and \cite{Battefeld2015}
for a discussion in a broader context). Note that, besides the Ekpyrotic
solution, there are also other possibilities to isotropize the universe,
e.g., by adding non-linear effects in the matter content at high energy
scales~\cite{Bozza2009} or through quantum effects~\cite{Pinto-Neto2000}.
However, we do not think it is necessary at this stage to add such
effects to our models. Without them, and in a model already containing a
radiation fluid from the beginning, there is no need to have some sort of
reheating after the bounce in our models. Hence the present work
differentiates substantially from previous work on this subject
\cite{Quintin2014, Haro2015, Hipolito-Ricaldi2016}, where the bounce
mechanism is completely different, and an ekpyrotic phase is present. It
is worth mentioning that such approach, where one assumes an initial
contracting phase already homogeneous and isotropic containing no exotic
components, were previously study in many papers~\cite{Peter2003,
Pinto-Neto2005, Peter2005, Peter2006, Peter2007, Peter2008, Bessada2012}.

The paper is organized in the following way: in the next section
(Sec.~\ref{sec:parprod}), we present the essential properties of the
bouncing models we are considering, and the relevant quantities for the
particle creation computation. In Sec.~\ref{sec:radiation}, we calculate
the production of particles in bouncing models with a single radiation
fluid, for the conformal and minimal coupling case, and for massive and
massless scalar fields. In Sec.~\ref{sec:twofluids}, we will perform the
same calculations for bouncing models with two fluids, dust and
radiation. We conclude in Sec.~\ref{conclusion} with a summary of our
results, and a discussion concerning its perspectives. All numerical
calculations performed in this work were done using the Action Angle (AA)
variables method as described in Appendix~\ref{app:AAvar}. Moreover, we
also used the Wentzel-Kramers-Brillouin (WKB) approach, described in
Appendix~\ref{app:WKBM}. Both methods gave the same results within the
required precision.

\section{Particle Production in Bouncing Models}
\label{sec:parprod}

We will consider bouncing models where the physical effect which avoid
the usual classical cosmological singularity are quantum corrections to
the classical Friedmann equations. These quantum effects are calculated
from a Wheeler-DeWitt quantization of the background and interpreting
the solution using the de Broglie-Bohm quantum theory, see
Ref.~\cite{Pinto-Neto2013} and references therein.

For a single hydrodynamical fluid with $p=\lambda \rho$, the
Wheeler-DeWitt equation reads
\begin{equation}
i\frac{\partial\Psi_{(0)}(a,T)}{\partial T}= \frac{1}{4}
\frac{\partial^2\Psi_{(0)}(a,T)}{\partial \chi^2}, \label{es202}
\end{equation}
where we have defined
$$\chi=\frac{2}{3} (1-\lambda)^{-1} a^{3(1-\lambda)/2},$$
$a$ is the scale factor, and $T$ is the degree of freedom associated to
the fluid, which plays the role of time.

This is just the time reversed Schr\"odinger equation for a one
dimensional free particle constrained to the positive axis. As $a$ and
$\chi$ are positive, solutions which have unitary evolution must satisfy
the condition
\begin{equation}
\label{cond27} \Psi_{(0)}^{\star}\frac{\partial\Psi_{(0)}}{\partial
\chi}\Biggl|_{\chi=0} -\Psi_{(0)}\frac{\partial\Psi_{(0)}^{\star}}{\partial
  \chi}\Biggl|_{\chi=0}=0
\end{equation}
(see Ref.~\cite{Alvarenga2002} for details). We will choose the initial
normalized wave function
\begin{equation}
\label{initial}
\Psi_{(0)}^{(\mathrm{init})}(\chi)=\biggl(\frac{8}{T_b\pi}\biggr)^{1/4}
\exp\left(-\frac{\chi^2}{T_b}\right) ,
\end{equation}
where $T_b$ is an arbitrary constant. The Gaussian
$\Psi_{(0)}^{(\mathrm{init})}$ satisfies condition~\eqref{cond27}.

Using the propagator procedure of Refs.~\cite{AcaciodeBarros1998,
Alvarenga2002}, we obtain the wave solution for all times in terms of
$a$:
\begin{widetext}
\begin{equation}\label{psi1t}
\Psi_{(0)}(a,T)=\left[\frac{8 T_b}{\pi\left(T^2+T_b^2\right)}
\right]^{1/4}
\exp\biggl[\frac{-4T_ba^{3(1-\lambda)}}{9(T^2+T_b^2)(1-\lambda)^2}\biggr]
\exp\left\{-i\left[\frac{4Ta^{3(1-\lambda)}}{9(T^2+T_b^2)(1-\lambda)^2}
+\frac{1}{2}\arctan\biggl(\frac{T_b}{T}\biggr)-\frac{\pi}{4}\right]\right\}.
\end{equation}
\end{widetext}

Due to the chosen factor ordering, the probability density $\rho(a,T)$
has a non trivial measure, and it is given by $\rho(a,T) =
a^{(1-3\lambda)/2} \left|\Psi_{(0)}(a,T)\right|^2$.  Its continuity
equation coming from Eq.~\eqref{es202} reads
\begin{equation}
\label{cont} \frac{\partial\rho}{\partial T}
-\frac{\partial}{\partial a}\biggl[\frac{a^{(3\lambda-1)}}{2}
\frac{\partial S}{\partial a}\rho\biggr]=0 ,
\end{equation}
which implies in the de Broglie--Bohm quantum theory that
\begin{equation}
\label{guidance}
\frac{\dd{}a}{\dd{}T} = -\frac{a^{(3\lambda-1)}}{2} \frac{\partial
S}{\partial a} ,
\end{equation}
in accordance with the classical relations $\dot{a} = \{a, H\}=
-a^{(3\lambda-1)} P_a/2$ and $P_a=\partial S/\partial a$.

Inserting the phase of \eqref{psi1t} into Eq.~\eqref{guidance}, we obtain
the Bohmian quantum trajectory for the scale factor:
\begin{equation}\label{at}
a(T) = a_b
\left[1+\left(\frac{T}{T_b}\right)^2\right]^\frac{1}{3(1-\lambda)} .
\end{equation}
Note that this solution has no singularities, and tends to the classical
solution when $T\rightarrow\pm\infty$. We are in the time gauge $N =
a^{3\lambda}$, thus $T$ is related to conformal time through
\begin{equation}
\label{jauge} N\dd T = a \dd \eta \quad \Longrightarrow \dd\eta =
\left[a(T)\right]^{3\lambda-1} \dd T.
\end{equation}
Although this solution describe a bounce for any choice of barotropic
fluid, in this work we are interested in a bouncing scenario without any
exotic component. Hence, the matter component with largest equation of
state that will dominate the evolution for small $a$ will be radiation.
Thus, choosing $\lambda = 1 / 3$, we obtain that $\chi = a$, while $T$ is
simply the conformal time, i.e., $T = \eta$.

The solution~\eqref{at} can be obtained for other initial wave functions
(see Ref.~\cite{Alvarenga2002}). In Ref.~\cite{Peter2016} it was shown
that defining the wave function at an arbitrary initial time also leads
to bouncing solutions. For the symmetric bounce given in Eq.~\eqref{at},
the quantum effect which causes the bounce is equivalent to adding to
the Friedmann equation a term corresponding to a negative energy density
of a stiff matter fluid. This same behavior has been obtained in other
quantum cosmological bounce scenarios, as in~\cite{Bergeron2014}, and in
Refs.~\cite{Ashtekar2006, Taveras2008} when the background fluid has a
dust-like equation of state.

A more elaborated and detailed model containing two fluids, dust and
radiation, can be found in Ref.~\cite{Pinto-Neto2005}. The model
parameters can be chosen such that the radiation fluid dominates during
the bounce, and the dust fluid dominates far from the bounce scale. The
Bohmian solution for the scale factor coming from the phase of the wave
solution of the corresponding Wheeler-DeWitt equation reads
\begin{equation}
\label{eq:rm:a}
a(\h) = a_{e} \left[ \left( \frac{\h}{\h_{*}} \right)^{2}  + 2 \frac{\h_{b}}{\h_{*}} \sqrt{1 + \left( \frac{\h}{\h_{b}} \right)^{2}} \right],
\end{equation}
where $a_{e}$ is the scale factor at matter-radiation equality, and
parameters $\eta_*$ and $\eta_b$ are related to the wave-function
parameters [similarly to the case of a single fluid where the spread of
the initial Gaussian distribution $T_b$ ends up being the bouncing time
scale in Eq.~\eqref{at}].

It is, nonetheless, more convenient to reparametrize the bounce solution
using observable related quantities. In this section, all quantities
calculated at a time when the scale factor has the same value as today
will be denoted by the subscript ${}_0$. Expanding Eq.~\eqref{eq:rm:a}
for large $\eta$ we obtain the Hubble function
\begin{equation}\label{HC2}
H^2 \approx \frac{4 a_e}{\eta_*^2} \left(\frac{1}{a^3} + \frac{a_e}{a^4}\right),
\end{equation}
from where we can readily identify the dimensionless density parameters today $\Omega_{m0} = \rho_{m0}/\rho_{\mathrm{crit}0}$ and $\Omega_{r0} = \rho_{r0}/\rho_{\mathrm{crit}0}$ as the coefficients of $(a_0/a)^3$ and $(a_0/a)^4$, respectively
\begin{equation}\label{eq:dmless}
\Omega_{m0} = \frac{a_e}{a_0}\frac{4R_H^2}{\eta_*^2}, \qquad \Omega_{r0} = \left(\frac{a_e}{a_0}\right)^2\frac{4R_H^2}{\eta_*^2},
\end{equation}
where $\rho_{\mathrm{crit}0} = 3H_0^2/(8\pi{}G)$ is the critical density
today, $R_H \equiv 1/(a_0H_0)$ is the co-moving Hubble radius,
$\rho_{m0}$ and $\rho_{r0}$ the energy densities of matter and
radiation. Next, expanding the Hubble function for large $\eta_*$, i.e.,
considering the dust domination only in the far past, we get
\begin{equation*}
H^2 \approx H_0^2\Omega_{r0}\left(x^4 - \frac{\eta_b^2x^6}{R_H^2}\right),
\end{equation*}
where we introduced the redshift-like variable $x \equiv a_0 / a$. From
the expression above, we see that near the quantum bounce the Hubble
function evolves as a classical Hubble function in the presence of a
radiation fluid with density parameter $\Omega_{r0}$, and a stiff matter
fluid with negative density parameter given by
\begin{equation}\label{eq:OmegaQ}
\Omega_{q0} = - \frac{\Omega_{r0}}{x_b^2}, \qquad x_b \equiv \frac{R_H}{\eta_b\sqrt{\Omega_{r0}}}.
\end{equation}
Hence, the quantum effect we have calculated, which stops the
contraction and realizes the bounce, is dynamically equivalent to a
bounce caused by the presence of an additional stiff matter fluid with
negative energy, besides the usual matter and radiation fluids, in a
classical cosmological scenario obeying the Friedmann equation.
Note, however, that this equivalence is valid only at the background
level.

Using these new parameters, we obtain
\begin{equation}\label{eq:HQ2}
H^2 \approx H_0^2\Omega_{r0}x^4\left[1 - \left(\frac{x}{x_b}\right)^2\right].
\end{equation}
Consequently, $x_b$ provides the scale factor where the bounce takes
place (apart from a small correction coming from the dust matter
density). Finally, we can invert the expression above to obtain all
wave-function parameters in terms of the new observable related ones,
\begin{equation}
a_e = a_0 \frac{\Omega_{r0}}{\Omega_{m0}}, \quad \eta_* = 2R_H\frac{\sqrt{\Omega_{r0}}}{\Omega_{m0}}, \quad \eta_b = \frac{R_H}{x_b\sqrt{\Omega_{r0}}}.
\end{equation}


The curvature scale at the bounce can be calculated as
\begin{align*}
\label{curvature-scale}
L_b &= \left.\frac{1}{\sqrt{R}}\right\vert_{\eta=0} = \left.\sqrt{\frac{a^3(\eta)}{6a''(\eta)}}\right\vert_{\eta=0} \\
&= \frac{a_b \eta_b}{\sqrt{6\left(2\gamma_b+1\right)}} = \frac{1}{\sqrt{\left(2\gamma_b + 1\right)}}\frac{a_0R_H}{x_b^2\sqrt{6\Omega_{r0}}},
\end{align*}
where $R$ is the four dimensional Ricci scalar and 
\begin{equation}\label{eq:gamma}
\gamma_b \equiv \frac{\Omega_{m0}}{(4x_b\Omega_{r0})},
\end{equation}
is the ratio of the dust and radiation matter density at the bounce
(where the factor of 4 was included for later convenience). Imposing
that the bounce scale is larger than the Planck scale, $L_b > L_\Pl$, we
can obtain an upper bound on $x_b$. This bound is relevant since we
expect that the Wheeler-DeWitt equation should be a valid approximation
for any fundamental quantum gravity theory only at scales not so close
to the Planck length. Using $H_0 = 70
\left[\mathrm{Km\;s^{-1}\;Mpc^{-1}}\right]$ we obtain
$$\frac{a_0R_H}{L_\Pl} \approx 8 \times 10^{60},$$
and consequently,
$$ 
x_b \lesssim \frac{\sqrt{8} 10^{30}}{(6\Omega_{r0})^{1/4}} \approx 2\times10^{31}.
$$
In the above calculation, we assumed $\gamma_b \ll 1$ because the bounce
must also happen at energy scales higher than the start of the
nucleosynthesis (around $10\;\mathrm{MeV}$), which implies $x_b \gg
10^{11}$ (using cosmic microwave background radiation temperature
$T_{\gamma0}\approx2.7\,\mathrm{K}$), and we are assuming that
$\Omega_{r0}$ should not be much smaller than its usual value
$\Omega_{r0} \approx 8 \times 10^{-5}$. Hence we get
\begin{equation}\label{eq:xb:range}
10^{11} \ll x_b < 2 \times 10^{31}.
\end{equation}

\begin{figure*}
\centering \includegraphics[scale=1.0]{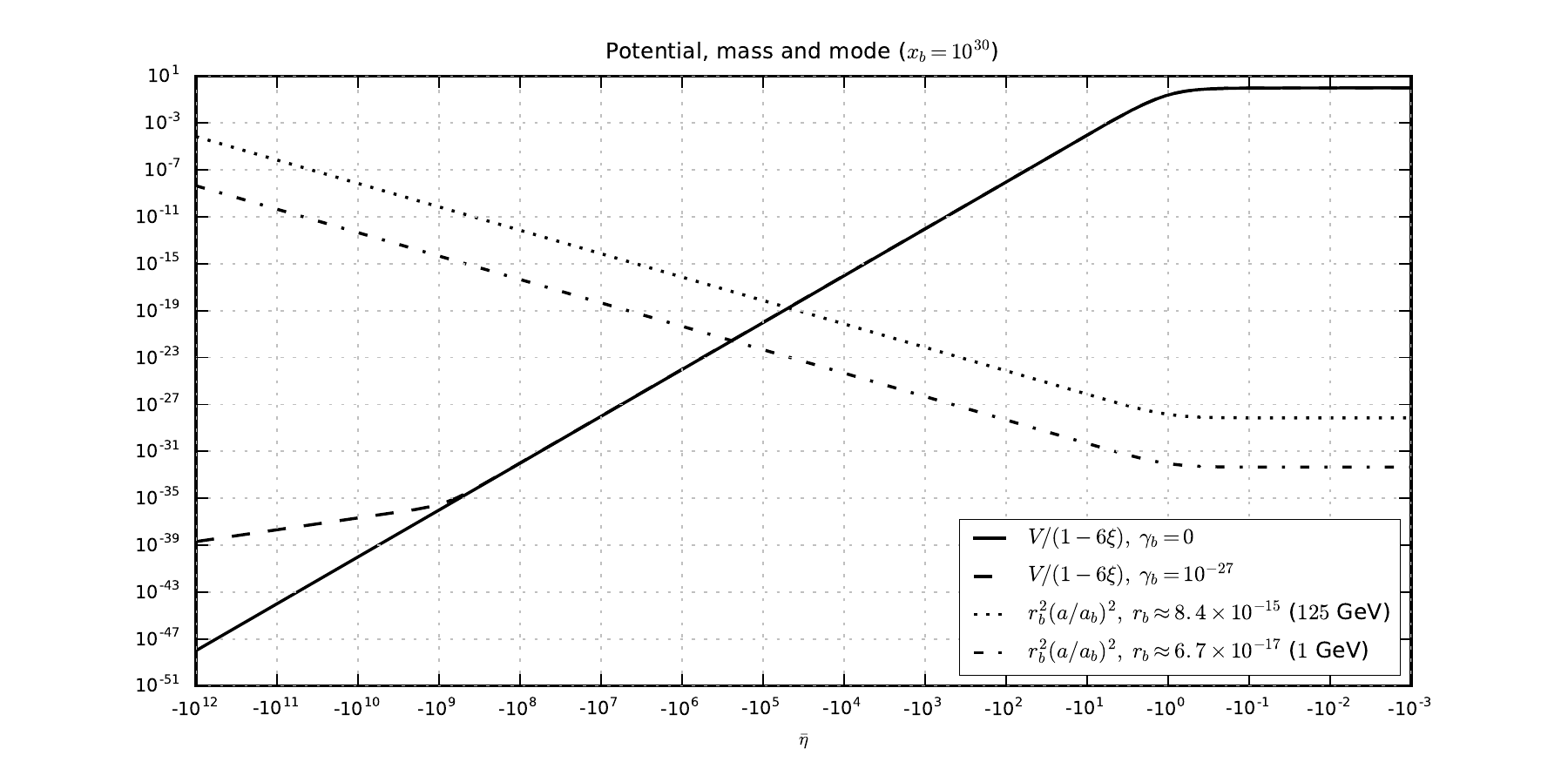} \caption{The
gravitational potential $V=a''/a$ and the mass term $m^2 a^2$ for the
parameters discussed in Sec.~\ref{sec:parprod}. In the initial phase,
the potential $V$ grows as a power law. If there is dust, then the
power-law changes during the dust radiation transition. The potential
attains its maximum near the bounce. In the minimally coupled case ($\xi
= 0$), the potential maximum is $1$. These features of the potential $V$
are shown in the continuum and dashed lines. In the massive case, the
mass term dominates at early times, larger the mass longer the time
interval it dominates the mode evolution. This is shown by the dotted
and dot-dashed lines in the figure. Note that the mass term
dominates the gravitational potential $V$ up to the radiation dominated
epoch, unless the particle mass is very small ($m < 10 \,\mathrm{eV}$).
Hence the presence of dust does not affect much the particle production.
For the same reason, the solutions at past and future infinity do not
depend on $\xi$, unless it becomes very close to the conformal coupling.
} \label{fig:RadScalarsPotential}
\end{figure*}

Using the parameters above, we make the definitions
\begin{equation*}
\label{TimeScale0}
\bar{\eta} = \frac{\h}{\h_{b}}, \quad \bar{k} = k \h_{b}, \quad \text{and} \quad r_b = m a_{b} \h_{b},
\end{equation*}
which are the natural parameters appearing in the equations we will
solve, as we will see in the following sections. With this definition, it
is easy to see that
\begin{equation}
\label{rb}
r_b = \frac{a_b\eta_b}{L_C} \approx \frac{L_b}{L_C}, \qquad L_C \equiv \frac{1}{m},
\end{equation}
where $L_C$ is the Compton wavelength of the massive particle. Note that
usually $r_b\ll 1$ because the curvature scale at the bounce is much
smaller than the Compton wavelength, or the mass of the particle is much
smaller than the mass-energy scale at the bounce.

In terms of the new parameters we write the Bohmian trajectory as
\begin{equation}\label{atnp}
a(\eta) = a_b\left(\bar{\eta}^2\gamma_b + \sqrt{1+\bar{\eta}^2}\right),
\end{equation}
Note that, the dust and radiation terms have equal weight at
$\bar{\eta}_e \approx 1/\gamma_b$, which is the same result one would
obtain substituting $a_e$ in the equation above. In the case of pure
radiation ($\Omega_{m0} = 0$ and, therefore, $\gamma_b = 0$), the scale
factor reduces to
\begin{equation} \label{eq:aR}
a(\h) = a_{b} \sqrt{1 + \bar{\eta}^2},
\end{equation}
which is exactly the trajectory given in Eq.~\eqref{at}.

The Klein-Gordon equation for the modes of a free massive scalar field $\varphi$ in a Friedmann background reads \cite{Birrell1982}
\begin{equation} \label{eq:KG}
\frac{\dd^2 \phi_{k} (\eta)}{\dd\eta^{2}} + \left[ k^{2} + m^{2} a(\eta)^2 - \frac{\left(1 - 6\xi\right)}{a(\eta)}\frac{\dd^2a}{\dd\eta^2} \right] {\f}_k (\eta) = 0,
\end{equation}
where ${\f}_k \equiv a \varphi _k$. Using the scale factor from
Eq.~\eqref{atnp} and the natural parameters, the mode equation reads
\begin{align}\label{eom}
&\phi_k^{\prime\prime} + \left(\nu^2 - V \right) {\f}_k = 0, \\ \label{V}
&V \equiv \left(1 - 6\xi\right)\frac{[(1+\bar{\eta}^2)^{-3/2}+2\gamma_b]}{\sqrt{1+\bar{\eta}^2} + \bar{\eta}^2\gamma_b},\\
&\nu^2 \equiv \bar{k}^{2} + r_b^2 \frac{a^2}{a^2_b},
\end{align}
where the prime denotes derivatives with respect to $\bar{\eta}$, i.e.,
${}^\prime \equiv \dd/\dd\bar{\eta}$, $V$ represent the potential felt by
the modes $\phi_k$, and $\nu$ the frequency of the mode $k$. It should be
pointed out that, in the presence of the dust fluid, the potential
changes twice: as one can see from Eq.~\eqref{V}, the radiation fluid
becomes dominant in the denominator at a different time than in the
numerator (at $\bar{\eta} \approx -1/\gamma_b$, and $\bar{\eta} \approx
-1/\gamma_b^{3/2}$, respectively, see also
Fig.~\ref{fig:RadScalarsPotential}). In the Appendix~\ref{app:AAvar} we
also realize that the function $\nu^\prime/\nu^2$ also plays the role of
a potential for $\phi_k$. Using the definitions above we get
\begin{equation}\label{eq:nupnu}
\frac{\nu^\prime}{\nu} = r_b^2\bar{\eta} \frac{\left[\left(1 + \bar{\eta}^2\right)^{-1/2} + 2\gamma_b\right]\left(\bar{\eta}^2\gamma_b + \sqrt{1+\bar{\eta}^2}\right)}{\bar{k}^2 + r_b^2\left(\bar{\eta}^2\gamma_b + \sqrt{1+\bar{\eta}^2}\right)^{2}}.
\end{equation}

Given a complete set of solutions for the mode equation Eq.~\eqref{eom},
a set of creation and annihilation operators is defined, and
consequently a vacuum state~\cite{Wald1994, Birrell1982, Chung2003}. The
ambiguity in defining the vacuum state stems from the fact that we do
not have a general procedure to define a unique set of modes when
space-time does not possess a global time-like Killing vector. One
special and suitable choice is the so called adiabatic vacuum
\cite{Birrell1982,Parker2009,Fulling1989}. One of the main physical
properties of this vacuum state choice is that its vacuum expectation
value of the number operator varies minimally when the expansion rate of
the universe becomes arbitrarily slow (see also Ref.~\cite{Chung2003}
for a good review on that). As discussed in Ref.~\cite{Chung2003}, for a
given mode $k$ at a time $\eta$, the adiabatic vacuum can be defined up
to a maximum order $N_{k,\eta}$.\footnote{This is a consequence of the
fact that the adiabatic expansion is asymptotic. However, in some
special cases the series is convergent, and the vacuum can be defined up
to an arbitrary function that decreases faster than any finite power of
$k^{-1}$.} The maximum order $N_{k,\eta}$ is a monotonically increasing
function of $k$. Therefore, large $k$'s have less ambiguity in their
vacuum definition than small $k$'s.

The adiabatic vacuum state has two points relevant to our problem.
First, it may depend on the time chosen to define it. If we impose the
adiabatic vacuum condition at a time $\eta_i$ and evolve the modes
through Eq.~\eqref{eom} until $\eta_f$, we may obtain a different set of
mode functions we would otherwise get by imposing the adiabatic vacuum
condition at $\eta_f$. The other point about this procedure, which is a
consequence of the existence of the maximum adiabatic order $N_{k,\eta}$
discussed above, is that it cannot be applied for all modes $k$, as it
depends on the behavior of the mode functions, and for a given time
$\eta$ only a subset of modes behave in an adiabatic
manner.\footnote{Alternatively, one can impose a vacuum state by
choosing a boundary condition. This asymptotic state is sometimes called
``Bunch-Davies vacuum''. Nonetheless, this choice is not free from
ambiguities, and coincides with the adiabatic vacuum up to its
approximation order~\cite{Chung2003}.}

The first point can be laid down as follows: given a mode $k$ at
$\eta_i$, we impose that the mode initial conditions for $\phi_k$ are
given by the adiabatic approximation up to the maximum order
$N_{k,\eta_i}$. Using this as initial condition on Eq.~\eqref{eom}, we
obtain the solution $\phi^{(i)}_k (\eta)$. Repeating the process at a
time $\eta_f$ and comparing both solutions at $\eta_f$, we have
\begin{equation}
\phi^{(i)}_k(\eta_f) - \phi^{(f)}_k(\eta_f) \lesssim \mathcal{O}\left(k^{-N_{k,[\eta_i,\eta_f]}}\right),
\end{equation}
where $N_{k,[\eta_i,\eta_f]}$ is the maximum adiabatic order attainable
in the interval $[\eta_i, \eta_f]$.\footnote{See for example Eq.~(33)
of~\cite{Chung2003}.} To measure the difference between vacua, we
introduce the norm squared of the Bogoliubov coefficients given by
\cite{Birrell1982, Parker2009}
\begin{equation} \label{bogocoef}
\left\vert\b^{(i,f)}_{k}\right\vert^{2} = \left| \f^{(i)}_{k} {\f^{(f)\prime}_{k}}  - \f^{(f)}_{k} {\f^{(i)\prime}_{k}} \right|^{2} \equiv n_{k}^{(i,f)},
\end{equation}
The quantity $n_{k}^{(i,f)}$ is the number density of particles with
mode $k$ measured by observers in the adiabatic vacuum defined at
$\eta_f$ if the initial state was the adiabatic vacuum defined at
$\eta_i$. 

Suppose that the adiabatic approximation is valid through the whole
interval $[\eta_i, \eta_f]$. It means that there is some
$N_{k,[\eta_i,\eta_f]} > 0$, and consequently
\begin{equation}
n_{k}^{(i,f)} \lesssim \mathcal{O}\left(k^{-N_{k,[\eta_i,\eta_f]}}\right).
\end{equation}
In other words, in the ultraviolet limit (UV) $k\to\infty$, the
$\bar{k}^2$ frequency present in Eq.~\eqref{eom} dominates over all other
terms. Also, it can be shown that the maximum adiabatic order
$N_{k,\eta}$ increases to infinity in this limit. This is equivalent to
say that in the ultraviolet limit there is a strong suppression in the
number of particles created: as the adiabatic order goes to infinity, any
ambiguity in the vacuum definition must fall faster than any finite power
of $k^{-1}$, as an exponential decay. Hence, there is no divergence in
the UV limit, and the particle production is finite (unless some infrared
divergence is present).

Inspecting Fig.~\ref{fig:RadScalarsPotential}, we note that the
potential $V$ has a maximum at the bounce, $V(\eta=0) \approx 1 - 6\xi$.
Thus, any mode with $\bar{k} \gg \vert 1 - 6\xi\vert$ will be in the
adiabatic regime during its whole evolution, including through the
bounce itself, and hence particle production of such modes will be
exponentially suppressed. This was verified numerically, as we will see.
In other words, the maximum of the potential provides a natural cutoff:
any mode $\bar{k}^2$ larger than it will be strongly suppressed in the
particle creation. On the other hand, for modes smaller than the
potential maximum, there is a time interval where the adiabatic
approximation fails, and particles will be produced.

An important comment has to be made now: in Ref.~\cite{Quintin2014}, all
calculations are done for modes much less than the Hubble radius at all
epochs of their cosmological scenario, which is physically the
ultraviolet limit ($\bar{k}^2 \gg V$). Hence, as we discussed above,
there is an exponential cutoff for these modes, and particle production
is heavily suppressed. Another way to phrase it can be: for modes which
never cross the potential, the adiabatic vacuum solution, which matches
the boundary condition in the far past before the bounce, is always a
good approximation at any time. Hence, it coincides with the solution
obtained through the adiabatic boundary condition prescribed in the far
future after the bounce. Consequently, one must have $\beta_k \approx 0$
for these modes. The particle production which is obtained in
Ref.~\cite{Quintin2014} comes from the matching conditions they impose,
which does not capture the precise quantitative evolution of mode
function for these large $k$ modes. In other words, they artificially
introduce a background discontinuity through the matching approximation.
Note that, in Ref.~\cite{Haro2015}, a discontinuity in the background
between the different phases is assumed from the beginning, and this
gives rise to particle production which depends explicitly on the assumed
discontinuity. In practice, the discontinuity creates an infinite
potential, invalidating the adiabatic approximation at that point.

As for the second point, we want to calculate the amount of particle
creation using the above adiabatic vacuum prescription and for all modes
$k$. Thus, if the adiabatic vacua are defined at $\bar{\eta}_i
\rightarrow -\infty$ and $\bar{\eta}_f \rightarrow \infty$, the first
order adiabatic vacuum state, given by the zeroth order WKB solution of
equation Eq.~\eqref{eom}, coincides with the infinite order adiabatic
vacua, and hence they precisely define state solutions for all modes $k$.
In practice, we will consider that the scalar field is initially in the
adiabatic vacuum state in the far past $(\bar{\eta}_i \ll -1)$, compute
the evolution of such modes until the expansion era far from the bounce
($\bar{\eta}_f \gg 1$), and compare it to the scalar field solution at
$\bar{\eta}_f$ with boundary condition of being the adiabatic vacuum at
$\bar{\eta}_f$.  That is why we numerically prescribe our initial
conditions far from the bounce, in the past and in the future, through
first order adiabatic approximated solutions, and we verify that the
initial approximations coincide with the numerical solutions obtained
with such boundary conditions for a long interval of time $\bar{\eta}$
before the adiabatic approximation looses its validity.

We are also interested in the energy density of the created particles.
Using a vacuum definition at $\eta_i$ as our system state, the expected
value of the energy density at any time $\eta$ with respect to this
vacuum is
\begin{align} \nonumber
\left\langle\rho\right\rangle_{(i)} =
\frac{1}{a^4\eta_b}\int\!\!\frac{\dd^{3}k}{(2\p)^3} \frac{1}{2} \Bigg[&\left\vert{\f}_k^{(i)\prime}\right\vert^2 + \left(\nu^2 + \mathcal{H}^2\right)\left\vert{\f}_k^{(i)}\right\vert^2\\ \label{energy-density}
&- \mathcal{H}\left({\f}_k^{(i)*}{\f}_k^{(i)}\right)^\prime\Bigg],
\end{align}
where $\mathcal{H} \equiv a^\prime/a$.

As it is well known, the energy density of a scalar field in curved space
time is divergent. Not only the usual divergence obtained in Minkowsky
but new ones must be taken care of. Nonetheless, in this work we want the
study the amount of energy resulting from the particle creation during
the bounce phase. For this reason we introduce the expected value of the
energy density for the adiabatic vacuum defined at $+\infty$, i.e.,
$\left\langle\rho\right\rangle_{(f)}$. Then, the difference
\begin{equation}
\Delta\rho = \left\langle\rho\right\rangle_{(i)} - \left\langle\rho\right\rangle_{(f)},
\end{equation}
given us a finite quantity (see~\cite{Birrell1982}) representing the amount of energy created by the bounce phase. We can relate the mode functions associated
to both states as
\begin{equation}
\phi^{(f)}_k(\bar{\eta}) = \alpha_k^{(i,f)} \phi^{(i)}_k(\bar{\eta}) + \beta_k^{(i,f)} \phi^{(i)*}_k(\bar{\eta}).
\end{equation}
The constants $\alpha_k^{(i,f)}$ and $\beta_k^{(i,f)}$ can be readily calculated using Eqs.~\eqref{eq:alpha} and \eqref{eq:beta}. In the far future, when the modes $\phi_k^{(f)}$ are deep in the adiabatic phase, the energy difference is given by
\begin{equation}
\begin{split}
\Delta\rho =
\frac{1}{2\pi^2\eta_b^4a^4}\int_0^\infty\!\!\dd\bar{k}\;\bar{k}^2n_{k}^{(i,f)}\nu,
\end{split}
\end{equation}
and the number density of created particles is
\begin{equation} \label{eq:n}
n = \frac{1}{2\p ^2 \eta_b^3a^3} \int_0^\infty \dd \bar{k}\; \bar{k}^2 n_{k}^{(i,f)}.
\end{equation}
In the massive case, this provides the energy density when $a$ is large enough and consequently $\nu \approx a r_b$ for modes relevant to the integral above, i.e., when modes satisfying $\bar{k} \ll a r_b /a_b$ are dominant for the integral. As we will see below, the particle number density $n_{k}^{(i,f)}$ has an exponential cutoff in the ultraviolet limit, therefore, for the massive case and $a$ large enough, we have
\begin{equation}\label{eq:massive}
\begin{split}
\Delta\rho \approx m n,
\end{split}
\end{equation}
In the massless case, $m = 0$, where the frequency term is $\nu =
\bar{k}$, the energy density yields the usual result for a relativistic
fluid,
\begin{equation} \label{eq:rad}
\Delta\rho = \frac{1}{2\pi^2 \eta_b^4a^4}\int_0^\infty \dd\bar{k}\;\bar{k}^3n_{k}^{(i,f)}.
\end{equation}

\begin{figure*}
\centering \includegraphics[scale=0.45]{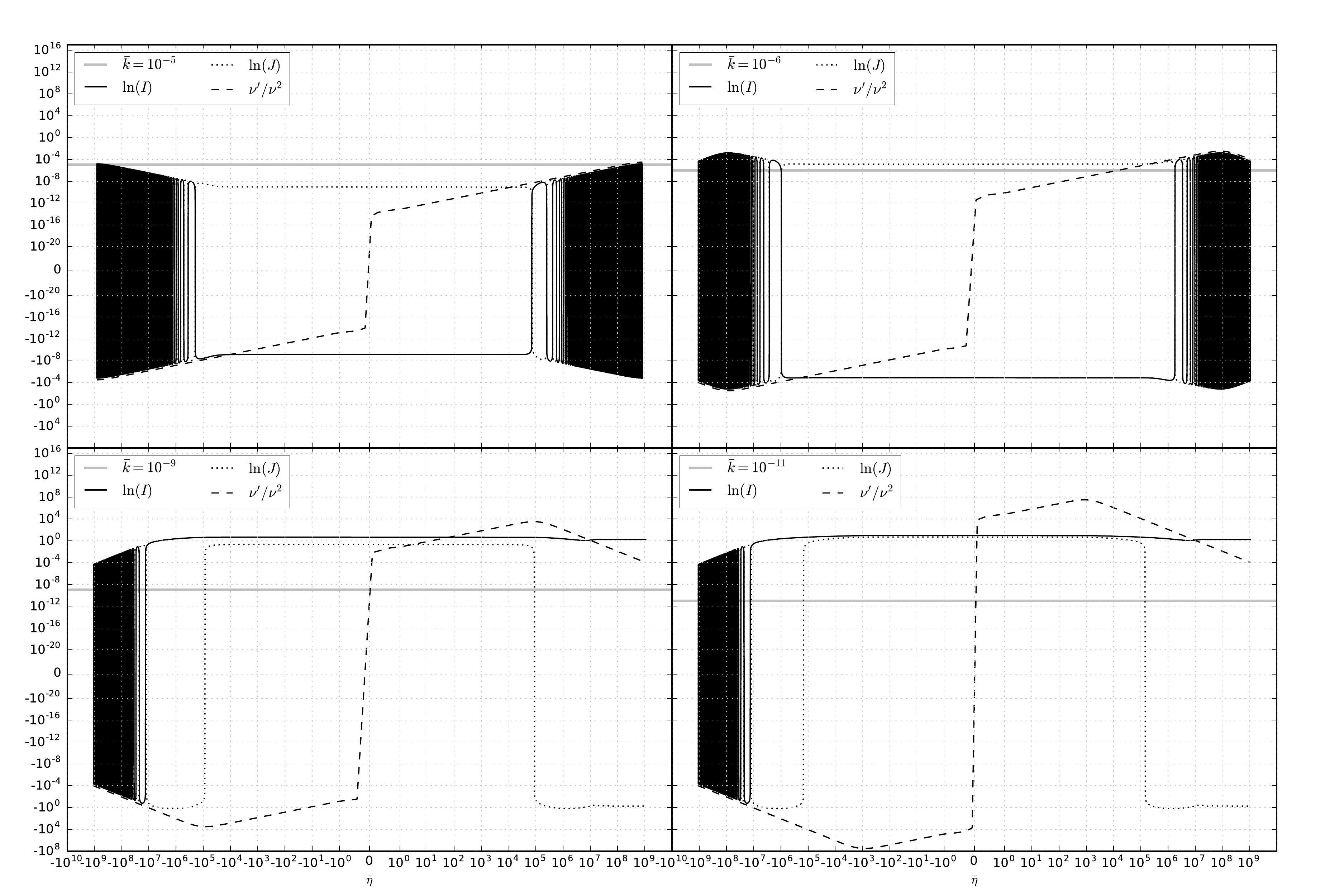}
\caption{Evolution of the adiabatic invariants for the conformal
coupling case using the Higgs mass $m_H$. The four figures show the
evolution for different values of $k$. The first one, in the top left
panel, shows a mode $k$ high enough such that the angle evolution never
gets dominated by the $\nu^\prime/\nu^2$ term. The oscillations, in this
case, never cease. The gap appearing in this panel is just the result of
using a logarithm scale in $\bar{\eta}$, otherwise, we would get regular
oscillations in $\bar{\eta}$. The top right panel shows the evolution
for a larger $k$. In this case, it becomes more evident the two
oscillatory phases appearing in
Eqs.~(\ref{eq:approx:I}--\ref{eq:approx:J}) which also show that the
amplitude of their evolutions is $\nu^\prime/\nu^2$. Consequently, any
change in the evolution of this factor can be seen in $I$ and $J$. The
lower left panel shows unequivocally that when the term
$\nu^\prime/\nu^2$ becomes larger than one, it dominates the evolution
of the adiabatic invariants, and stops their oscillatory behaviors. In
this case, the adiabatic invariants no longer oscillate after the bounce
and instead freeze out to constant values. We can also see that during
the $\nu^\prime / \nu^2$ dominated time interval, the adiabatic
invariants change very little, growing in the contracting phase and
shrinking back to one in the expanding phase, when the potential gets
smaller than one. Finally, in the lower right panel, we can see the same
behavior for a smaller value of $k$.} \label{fig:evol:cc}
\end{figure*}

\section{Bouncing with a Radiation Fluid}
\label{sec:radiation}

In this section, we consider a Bohmian solution of the Wheeler-DeWitt equation obtained in \cite{Peter2007} for the case of a universe dominated by a radiation perfect fluid only. In this case the scale factor is described by Eq.~\eqref{eq:aR}.

The Klein-Gordon equation Eq.~\eqref{eom} simplifies to
\begin{equation} \label{eq:KG:r}
\phi_k^{\prime\prime} + \left[\bar{k}^{2} + r_{b}^{2} \left( 1 + \bar{\eta}^{2} \right) - \frac{\left(1 - 6\x\right)}{\left(1 + \bar{\eta}^{2} \right)^{2}} \right] \phi_k = 0.
\end{equation}
Note that the $\x$ dependent term in~\eqref{eq:KG:r} goes to zero at
both past and future infinity, $\bar{\eta} \rightarrow \pm \infty$, for
whatever value of $\x$. Thus, the vacuum solutions at these asymptotic
limits do not depend on $\x$. The $\nu^\prime/\nu$ term
[Eq.~\eqref{eq:nupnu}] simplifies to
\begin{equation}\label{eq:nupnu:r}
\frac{\nu^\prime}{\nu} = \frac{r_b^2\bar{\eta}}{\bar{k}^2 + r_b^2(1 + \bar{\eta}^2)}.
\end{equation}
Comparing to $V/\nu$ this term also goes to zero at
$\left\vert\bar{\eta}\right\vert\to\infty$ but at a slower rate, i.e.,
$V / \nu \propto \bar{\eta}^{-5}$ and $\nu^\prime / \nu \propto
\bar{\eta}^{-1}$.

Figure \ref{fig:RadScalarsPotential} shows the gravitational potential
and the mass term for many different cases. The potential goes to zero
in two different ways, depending on the presence of dust matter. Thus,
in the massless case, the total modification to the frequency vanishes
asymptotically, giving simple plane-waves as asymptotic solutions
of~\eqref{eq:KG:r} for any value of $\x$. In the massive case, the mass
term dominates for $\vert\bar{\eta}\vert \gg 1$, while the $\x$
dependent term vanishes in this limit. Hence, the asymptotic solutions
are mass dependent, but still do not depend on $\x$.

\subsection{Conformal Coupling}

Let us start with the simpler conformally coupled case ($\x = 1/6$),
which is well-known in the literature \cite{Audretsch1978,
Audretsch1978a, Birrell1982}. The time dependent term of the frequency
is just the mass term, exemplified in Fig.~\ref{fig:RadScalarsPotential}
and Eq.~\eqref{eq:KG:r}, reduces to
\begin{equation} \label{eq:KG:r Conformal}
\phi_k^{\prime\prime} + \left[\bar{k}^{2} + r_{b}^{2} \left( 1 + \bar{\eta}^{2} \right) \right] \phi_k = 0.
\end{equation}
This equation has exact solutions in terms of parabolic cylinder
functions \cite{Gradshteyn2014}. The normalized solutions that match the
adiabatic vacuum solution are \cite{Birrell1982}
\begin{align} \label{eq:exact:sol}
 \f^{(i)}_{k} (\bar{\eta}) &=  \frac{e^{-\frac{\p}{8} \l }}{\left( 2 r_{b} \right)^{1/4}} D_{\frac{i\l -1}{2}} \left[ \left(i-1\right) \sqrt{r_{b}} \bar{\eta}\right], &\bar{\eta} < 0, \\
 \label{eq:exact:cc}
 \f^{out}_{k} (\bar{\eta}) &= \f^{in}_{k} (-\bar{\eta})^{*},  &\bar{\eta} > 0,
\end{align}
where
\begin{equation}
\l = r_{b}\left(1 + \frac{\bar{k}^{2}}{r_{b}^2}\right).
\end{equation}

To calculate the Bogoliubov coefficients, we use the identity \cite{Gradshteyn2014}
\begin{equation} \label{Parabolic Function Identity}
D_{\nu}(z) = e^{i \pi \nu} D_{\nu}(-z) + \frac{\sqrt{2\pi}}{\Gamma(-\nu)} e^{i \frac{\pi}{2}(\nu+1)} D_{-\nu-1}(-iz),
\end{equation}
to show that, for $\bar{\eta} > 0$,
\begin{equation}
\f^{(i)}_{k}(\bar{\eta}) = \frac{\sqrt{2\pi}e^{i \pi /4}}{\Gamma\left(\frac{1-i\lambda}{2}\right)} e^{-\pi \lambda /4} \f^{(f)}_{k}(\bar{\eta}) - i e^{-\pi \lambda /2} \f^{(f)*}_{k}(\bar{\eta}).
\end{equation}
It follows that
\begin{equation} \label{eq:BOGO:conformal}
n_{k}^{(i,f)} = e^{- \p \l} = \mbox{exp} \left[ -\p r_{b}\left(1  + \frac{\bar{k}^{2}}{r_{b}^2} \right)  \right],
\end{equation}
which falls off faster than any inverse finite power of $k$ or $m$. As it was explained in Ref.~\cite{Audretsch1978}, this is the spectrum of a non-relativistic thermal gas of particles with physical momentum $k/a$, chemical potential, and temperature, respectively given by
$$
\mu = - \frac{r_b }{2a_b\eta_b}\frac{a_b^2}{a^2}, \qquad T = \frac{1}{2\p a_b\h_{b} \kappa_{b} }\frac{a_b^2}{a^2},
$$ 
where $\kappa_{b}$ is the Boltzmann constant.

From Eq.~\eqref{eq:n}, the number density of created particles is
\begin{equation} \label{eq:pp:cc}
\begin{split}
n &= \frac{1}{2\p^2 a^{3}\eta_b^3} \int \dd\bar{k} \bar{k}^2 n_{k}^{(i,f)} = \left(\frac{\sqrt{r_b}}{2a\eta_b}\right)^3 \exp(-\pi r_b),\\
&= 
\left(\frac{\sqrt{\Omega_{r0}}}{a_0R_HL_C}\right)^{3/2}\frac{x^3}{8} e^{-\pi r_b}
\end{split}
\end{equation} 

The quantity $r_b = a_b\eta_b / L_c \approx L_b/L_C$ is usually very
small. For instance, assuming $x_b = 10^{30}$ (which gives roughly $L_b
\approx 10^3 L_\Pl$) and the Higgs particle, one of the most massive
scalar particles in the standard model, one gets $r_b = 8.4\times
10^{-15}$. Hence, we can neglect the exponential in Eq.~\eqref{eq:pp:cc}
due to the smallness of $r_b$. In other words, the exponential provides a
very large mass cutoff
$$
m_\mathrm{cutoff} = \left(\frac{x_b}{10^{30}}\right)^2\, 4.7\times 10^{15}\,\mathrm{GeV}.
$$  
Then, using Eq.~\eqref{eq:massive}, one can write the energy density of
created particles for very large scale factors as
\begin{equation}
\begin{split}
\Delta\rho = m n &\approx m \biggl[\frac{x(\Omega_r)^{1/4}}{2(R_H L_c)^{1/2}}\biggr]^3,\\
&\approx \biggl(\frac{m}{m_H}\biggr)^{5/2} x^3 \times 10^{-44}\; {\rm g/cm^3},
\end{split}
\end{equation} 
where $m_H$ is the Higgs mass ($125\,\mathrm{GeV}$),
yielding
\begin{equation}
\Omega_{\phi} \equiv \frac{\Delta\rho}{\rho_{\mathrm{crit}0}} \approx \biggl(\frac{m}{m_H}\biggr)^{5/2}x^3 \times 10^{-15}.
\end{equation}
The particle density does not depend on the bounce depth, at least not
for the cases considered, for masses closer or larger to the cutoff
$m_\mathrm{cutoff}$ the particle number density is further reduced by the
exponential factor.

\begin{figure*}
\centering \includegraphics[scale=0.45]{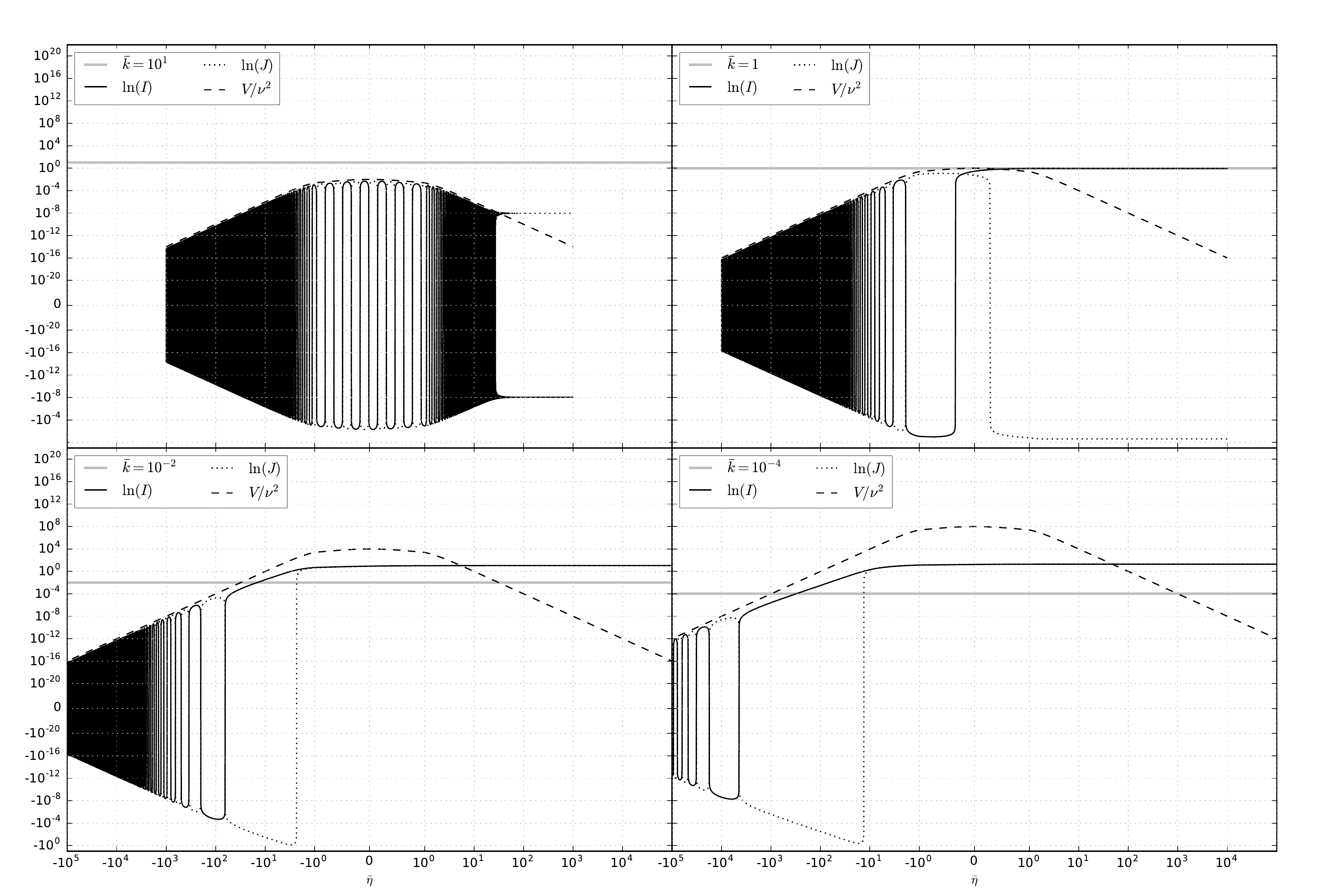}
\caption{Evolution of the adiabatic invariants for the minimal coupling
and massless case without dust. The four figures show the evolution for
different values of $k$. The description follows closely the discussion
of Fig.~\ref{fig:evol:cc}, the main difference being that in this case
the cutoff happens at $\bar{k} = 1$. Hence, in the sequence we see the
adiabatic regime valid during all times for $\bar{k} > 1$ and the
amplitude amplification of the adiabatic invariants as the value of
$\bar{k}$ decreases. } \label{fig:evol:min:massless}
\end{figure*}

We have also calculated Eq.~\eqref{eq:BOGO:conformal} numerically, using
the technique described in Appendix~\ref{app:AAvar}, in order to test
our algorithm. Later we will follow analogous steps to calculate the
Bogoliubov coefficients for the minimal coupling case. To do so, we have
used the definition of the Bogoliubov given in Eq.~\eqref{bogocoef} [and
in terms of AA variables by Eq.~\eqref{eq:beta2}]. Examining
Eqs.~\eqref{eq:theta} and \eqref{eq:I}, we see that $\nu^\prime/\nu^2$
controls the evolution. If $\nu^\prime/\nu^2$ is always smaller than
one, then the amplitude of the forced oscillations on $\ln(I)$ will also
be smaller than one. In that case the angle $\theta$ will evolve
accordingly to the frequency $\nu$. When this happens, the
adiabatic approximation will be valid during the bounce and, as
discussed above, the particle production will be heavily suppressed. In
Fig.~\ref{fig:evol:cc} we see two examples of modes, top panels, where
$\nu^\prime/\nu^2$ never grows larger than one and the adiabatic
approximation is valid everywhere. On the other hand, for modes having
$\nu^\prime/\nu^2 > 1$ for some time interval, the evolution of both the
angles and the adiabatic invariants are dominated by this term stopping
the oscillations and increasing the value of $I$. Hence, when
$\nu^\prime/\nu^2$ drops bellow one again, the further evolution of $I$
will only add a small negligible oscillatory term. This kind of
evolution can be seem in the lower panels of Fig.~\ref{fig:evol:cc}. To
compute the numerical evolution of these modes, we used the approximated
solutions (Eqs.~\ref{eq:approx:I}--\ref{eq:approx:psi}) using the
condition [Eq.~\eqref{eq:condition}] with $\epsilon = 10^{-4}$, i.e.,
$\vert\nu^\prime/\nu^2\vert < 10^{-4}$. After this point we compute the
numerical evolution.

It is worth noting that, from the discussion above, if
$\nu^\prime/\nu^2$ never grows larger than one, then the particle
production must be strongly suppressed for all modes $k$ where that
happens. It is easy to check that the maximum (in absolute value) of
$\nu^\prime/\nu^2$ reads [see Eq.~\eqref{eq:nupnu:r}]
\begin{equation}\label{eq:etam}
\left.\frac{\nu^\prime}{\nu^2}\right\vert_{\bar{\eta}^m} = \pm \frac{2r_b}{3\sqrt{3}(\bar{k}^2+r_b^2)},\qquad \bar{\eta}^m = \pm\sqrt{\frac{\bar{k}^2+r_b^2}{2r_b^2}},
\end{equation}
This maximum attains its largest value at $\bar{k} = 0$. If this value is smaller than one, i.e.,
\begin{equation*}
r_b > \frac{2}{3\sqrt{3}} \approx 0.385,
\end{equation*}
then the particle production is severely weakened. Note from 
Eq.~\eqref{eq:pp:cc} that the exponential cutoff is active when
\begin{equation}
r_b > \frac{1}{\pi} \approx 0.318.
\end{equation}
Hence, we can see that the particle creation cutoff can be approximately
deduced by examining the term $\nu^\prime/\nu^2$ alone.

In contrast, if the mass appearing in $\nu^\prime/\nu^2$ satisfies
\begin{equation*}
r_b < \frac{2}{3\sqrt{3}} \approx 0.385,
\end{equation*}
then all modes in the interval 
$$
\bar{k} < \sqrt{r_b}\sqrt{\frac{2}{3\sqrt{3}} - r_b},
$$
will have a period in their evolution where the term $\nu^\prime/\nu^2$
becomes important, and particle creation takes place. For example, in
Fig.~\ref{fig:evol:cc}, a Higgs mass was used ($r_b = 8.4 \times
10^{-15}$), so particles with $\bar{k} \lesssim 6\times10^{-8}$ (lower
panels) are produced, while particle creation with $\bar{k} \gtrsim
6\times 10^{-8}$ (upper panels) is suppressed, because the adiabatic
approximation is valid everywhere.

For the modes where particle creation occur, away from the cutoff, we can
also infer the amplitude. In this case, the $\nu^\prime/\nu^2$ term
controls most of the behavior of the AA variables.
Equation~\eqref{eq:evol:I:nu} describes the evolution of $I$ during this
period. In our case, the function $\nu$ is an even function of
$\bar{\eta}$, hence, if the adiabatic evolution stops at $-\bar{\eta}_0$
(here the subindex ${}_0$ denotes an arbitrary initial time), then it
will restart at $\bar{\eta}_0$. In this case, Eq.~\eqref{eq:evol:I:nu}
shows that, the adiabatic invariant will return to its original value at
the point where $\nu^\prime/\nu^2$ ceases to dominate its evolution,
i.e., $I(\bar{\eta}_0) = I(-\bar{\eta}_0)$, independently of $k$. Also,
this point is defined as $\nu^\prime/\nu^2 \approx 1$, for any $k$.
Hence, looking at Eq.~\eqref{eq:approx:I}, we see that at this point all
modes evolve back to the same amplitude $I$, and we conclude that the
amplitude of $n_k^{(i,f)}$ does not depend on $k$ for modes away from the
cutoff. Note that this is consistent with Eq.~\eqref{eq:BOGO:conformal},
i.e., for small $k$, $\beta_k^{(i,f)} \approx 1$.

Figure~\ref{fig:RadCoeffsConformal} compares our numerical solution
using Eq.~\eqref{eq:beta2} with the exact one given by
\eqref{eq:BOGO:conformal}. The comparison shows that our numerical
calculation predict correctly the particle production in the conformal
coupling case.

\subsection{Minimal Coupling}

Next, we calculate numerically the Bogoliubov coefficients for
Eq.~\eqref{eq:KG:r} in the minimal coupling case, $\x = 0$, the
generalization for any constant value of $\x$ being simple enough. As it
can be seen from Eq.~\eqref{eq:KG:r}, the solutions do not depend on
$\x$ at a sufficient time distance from the bounce,
$\vert\bar{\eta}\vert \gg 1$. However, in this case, we have to
distinguish the massive case from the massless case: the massless case
is trivial only in the conformal coupling case, where there is no
particle production [Eq.~\eqref{eq:KG:r} reduces to a collection of free
harmonic oscillators in this situation].

\subsubsection{Massless Case}
\label{sec:massless}

In the massless minimally coupled case, the Eq.~\eqref{eom} for the
modes reduces to
\begin{equation} \label{eq:KG:mc:ml}
\phi_k^{\prime\prime} + \left[ \bar{k}^{2} - \frac{1}{\left( 1 + \bar{\eta}^{2} \right)^{2}} \right] \phi_k = 0.
\end{equation}
It can be seen from the equation above that the minimal coupling term goes to zero at infinity, and the equation reduces to that of a simple harmonic oscillator. The vacuum mode solution is then given by
\begin{equation} \label{eq:KG:pw}
\f_{k} (\bar{\eta})  = \left( 2 \bar{k} \right)^{-1/2} \phantom{.} \exp \left( - i \bar{k} \bar{\eta} \right).
\end{equation} 
In practice, from \eqref{eq:KG:mc:ml}, the plane-wave vacuum solution can be consistently used for $\bar{k} \gg 1/(1+\bar{\eta}^2)$. The same conclusions arise from the inspection of condition~\eqref{eq:condition} and the approximated solutions Eqs.~(\ref{eq:approx:I}--\ref{eq:approx:psi}).

Following the discussion for the conformal coupling case, here, the controlling term is simply 
$$ 
\frac{V}{\nu^2} = \frac{1}{\bar{k}^2\left(1+\bar{\eta}^2\right)^2}.
$$
Its maximum with respect to the time variable is simply $\bar{\eta} = 0$.
Therefore, any $\bar{k} < 1$ will result in particle creation, i.e.,
there is an interval where $V/\nu^2 > 1$, and the adiabatic approximation
no longer holds. On the other hand, the modes $\bar{k} > 1$ are adiabatic
during the whole evolution, and any particle production is suppressed.
Thus, in this case, we expect an exponential cutoff at $\bar{k} \approx
1$.

The adiabatic regime ends at approximately 
$$
\eta_0 = - \sqrt{\frac{1}{\bar{k}} - 1},
$$
where we are considering only modes that produces particles, i.e.,
$\bar{k} < 1$. Them, using the solution for the $V/\nu$ dominated given
in Eq.~\eqref{eq:noaiab:I}, and evolving from $\eta_0$ to $-\eta_0$
results in
\begin{equation}\label{eq:approx:I:min}
\begin{split}
I &= I_0 \left[1 + \left(\bar{k}\vert\eta_0\vert + \tan^{-1}\vert\eta_0\vert\right)^2\right],\\
 &\approx I_0\left(\frac{\pi}{4\bar{k}^2}-\frac{2\pi}{3\sqrt{\bar{k}}} + 1\right).
\end{split}
\end{equation}
In this case $I_0$ is also equal for all modes. The end of the adiabatic
regime is determined by $V/\nu^2 \approx 1$, and the amplitude of $I$
during this regime is also determined by this same quantity. Thus, we
expect from the computation above that $n_k^{(i,f)} \approx \bar{k}^{-2}$
for $\bar{k} < 1$. In Fig.~\ref{fig:evol:min:massless} we can see clearly
that the frozen value of $I$ and $J$ increases when $\bar{k}$ gets
smaller, consistently with the discussion above.

\begin{figure}
\begin{center}
\includegraphics[scale=0.47]{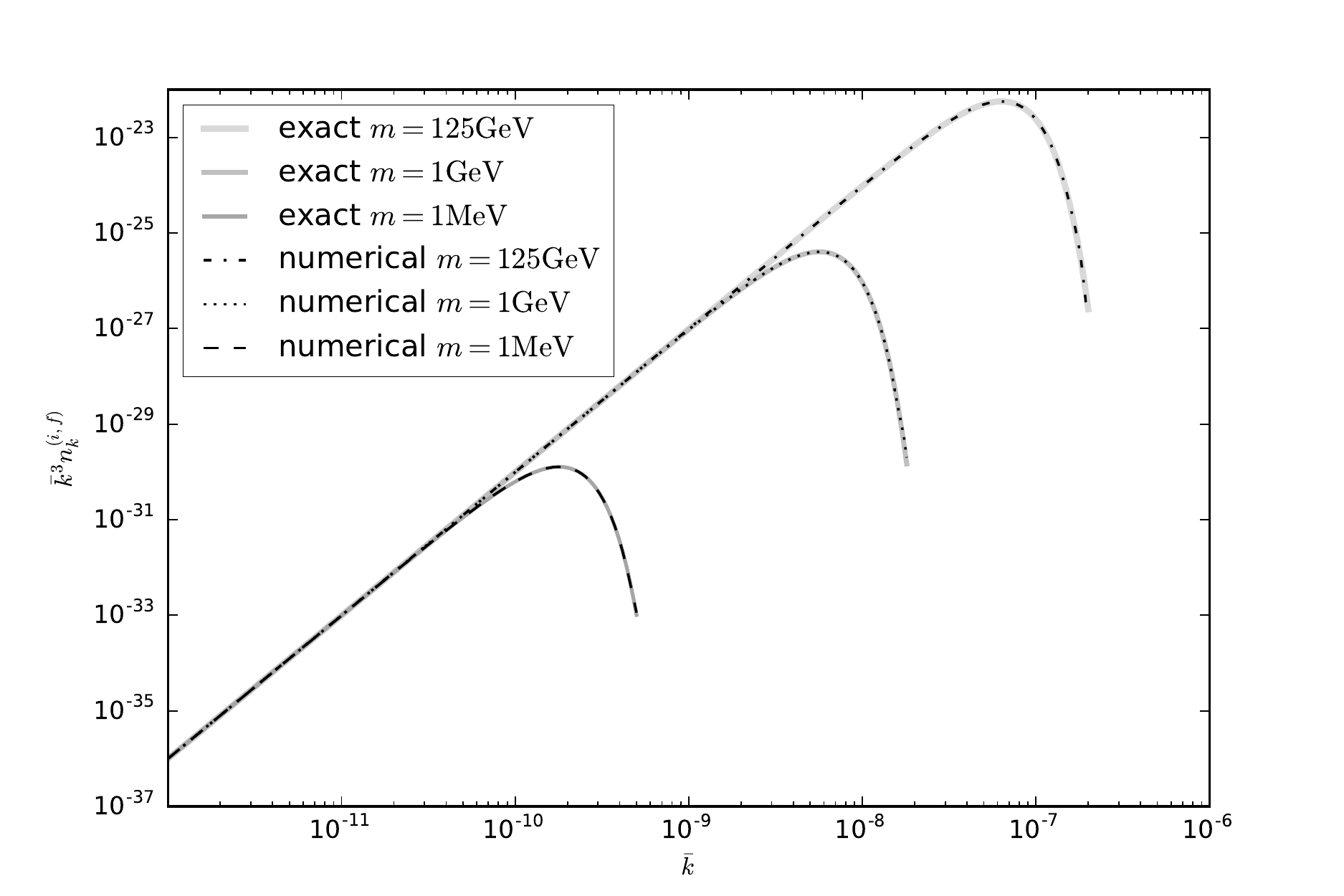} 
\end{center}
\caption{The Bogoliubov coefficients in the conformal coupling case with
the Higgs mass ($r_{b} = 8.4\times 10^{-15}$), $1\mathrm{GeV}$ mass
($r_b=6.7\times10^{-17}$) and $1\mathrm{MeV}$ mass
($r_b=6.7\times10^{-20}$). We compare our numerical solutions with the
exact solution given by Eq.~\eqref{eq:BOGO:conformal}. }
\label{fig:RadCoeffsConformal}
\end{figure}

We solved the full system using the AA variables and Eqs.~\ref{eq:theta},
\eqref{eq:I} and Eqs.~\eqref{eq:psi}, \eqref{eq:J},
(Eq.~\eqref{eq:KG:mc:ml}) numerically for $10^{-12} \leq \bar{k} \leq 10$
as described in Appendix~\ref{app:AAvar}, using $\epsilon = 10^{-4}$ to
obtain the initial integration time. As in the conformal coupling case,
for $\bar{k}$ greater than the maximal height of the potential, $\bar{k}
\gg 1$, the adiabatic approximation is valid everywhere and $n_k^{(i,f)}$
is negligible. In fact, as we discussed before, in this regime
$n_k^{(i,f)}$ decays exponentially with $\bar{k}$. For $\bar{k} \ll 1$,
one also could use the approximation methods developed in
Refs.~\cite{Mukhanov1992, Celani2011} to show our result above that
$\bar{k}^3n_k^{(i,f)} \approx {\pi^2 \bar{k}}/{4}$. We have verified
numerically these assertions. This means physically that only those modes
with wave number near the bouncing energy scale contribute effectively to
particle creation.

\begin{figure}
\begin{center}
\includegraphics[scale=0.47]{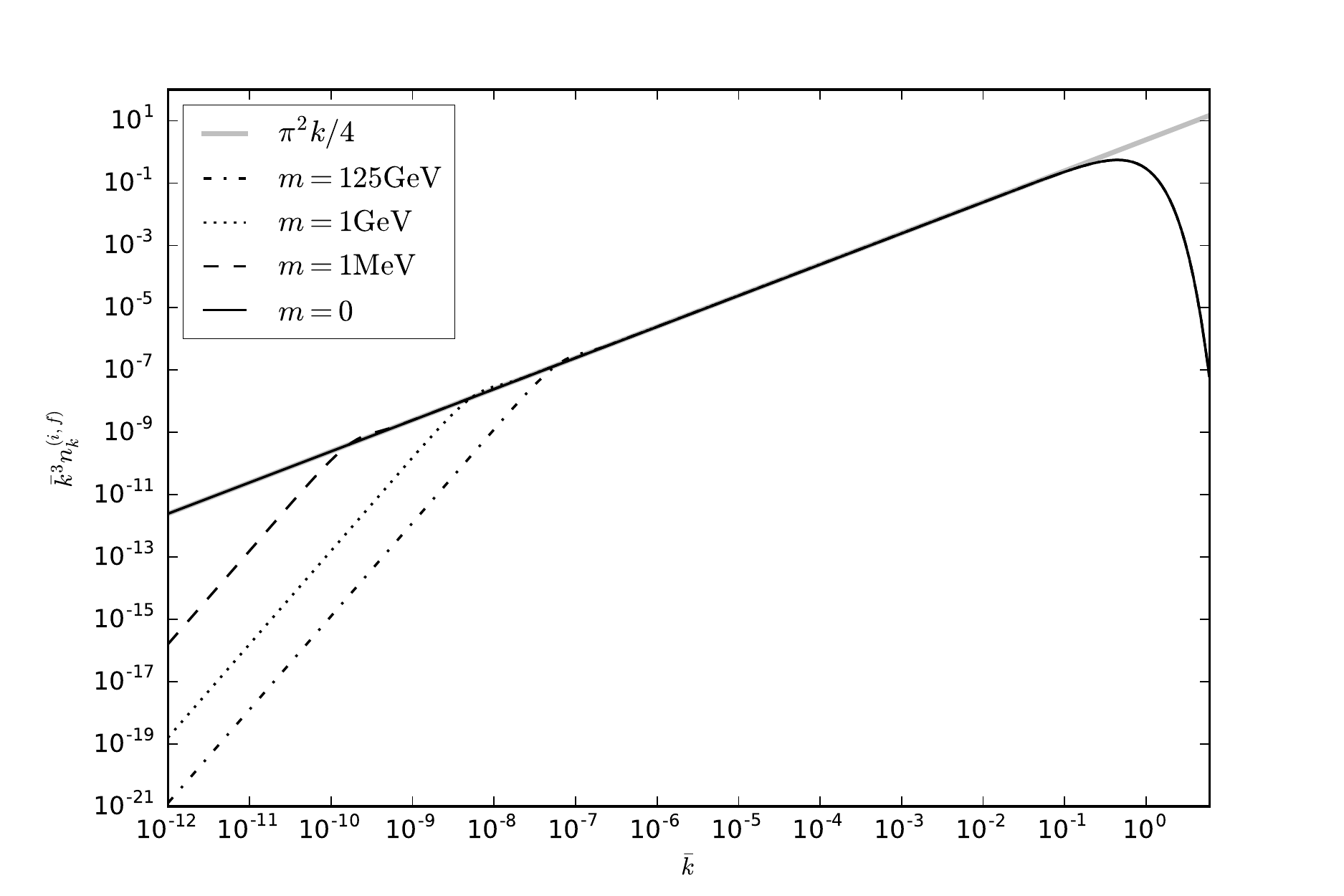} 
\end{center}
\caption{ The Bogoliubov coefficients in the minimal coupling case with
the Higgs mass ($r_{b} = 8.4\times 10^{-15}$), $1\mathrm{GeV}$ mass
($r_b=6.7\times10^{-17}$), $1\mathrm{MeV}$ mass
($r_b=6.7\times10^{-20}$) and the massless case. In this figure we also
plotted the first order approximation of the amplitude obtained in
Eq.~\eqref{eq:approx:I:min}, showing an excellent agreement with the
full numerical calculation. } \label{fig:RadCoeffsMinimal}
\end{figure}

To calculate the Bogoliubov coefficients, we use Eq.~\eqref{eq:beta2}
together with our numerical results. Figure~\ref{fig:RadCoeffsMinimal}
shows the Bogoliubov coefficients. Integrating numerically in $\bar{k}$,
we obtain the particle number density
\begin{equation}\label{eq:n:min}
n \approx \frac{6.7 \times 10^{-2}}{a^3\eta_b^3} \approx 3\times10^{-2}x^3\left(\frac{x_b}{10^{30}}\right)^3\;[\mathrm{cm}^{-3}].
\end{equation}
This result is the same for the massless and massive cases, within the
numerical precision. For massless particles, the energy density
calculated through Eq.~\eqref{eq:rad} gives
\begin{equation}\label{eq:rhoc:min}
\Delta\rho \approx \frac{3\times10^{-2}}{a^4\eta_b^4}, \quad \Omega_\phi = 2.5\times10^{-10}x^4\left(\frac{x_b}{10^{30}}\right)^4.
\end{equation}
Hence, only for $x_b > 10^{31}$ one should have a significant amount
of massless scalar particles, which could exceed the radiation density
in the universe. Note, however, that such $x_b$ would imply a $a_b\eta_b
> L_{\rm Pl}$, which goes beyond the scope of the calculations done in
this paper.

\subsubsection{Massive Case}

In the limit $\vert\bar{\eta}\vert \gg 1$, the coupling term depending on
$\x$ is negligible because it falls off with $\bar{\eta}^{-5}$ while the
mass related term goes with $\bar{\eta}^{-1}$. Thus, in this region, we
can use the exact solutions~\eqref{eq:exact:sol} and~\eqref{eq:exact:cc},
being understood that they are only approximately valid away from the
bounce. In the numerical calculation we used both the first order
adiabatic approximation and the exact solution to provide the initial
conditions for the numerical system (again, as described at
Appendix~\ref{app:AAvar}). The results are the same, within the required
precision. In order to test the consistency of the choice of initial
time, we evaluated the initial exact solutions along with our numerical
solutions. We did this for a number of values of the parameters, with
$\bar{k} \lesssim 10$ and $r_{b} \ll 1$, and found no inconsistency: the
numerical evolution was indistinguishable from the exact
solution~\eqref{eq:exact:sol} for sufficient early times.

\begin{figure*}
\centering \includegraphics[scale=0.45]{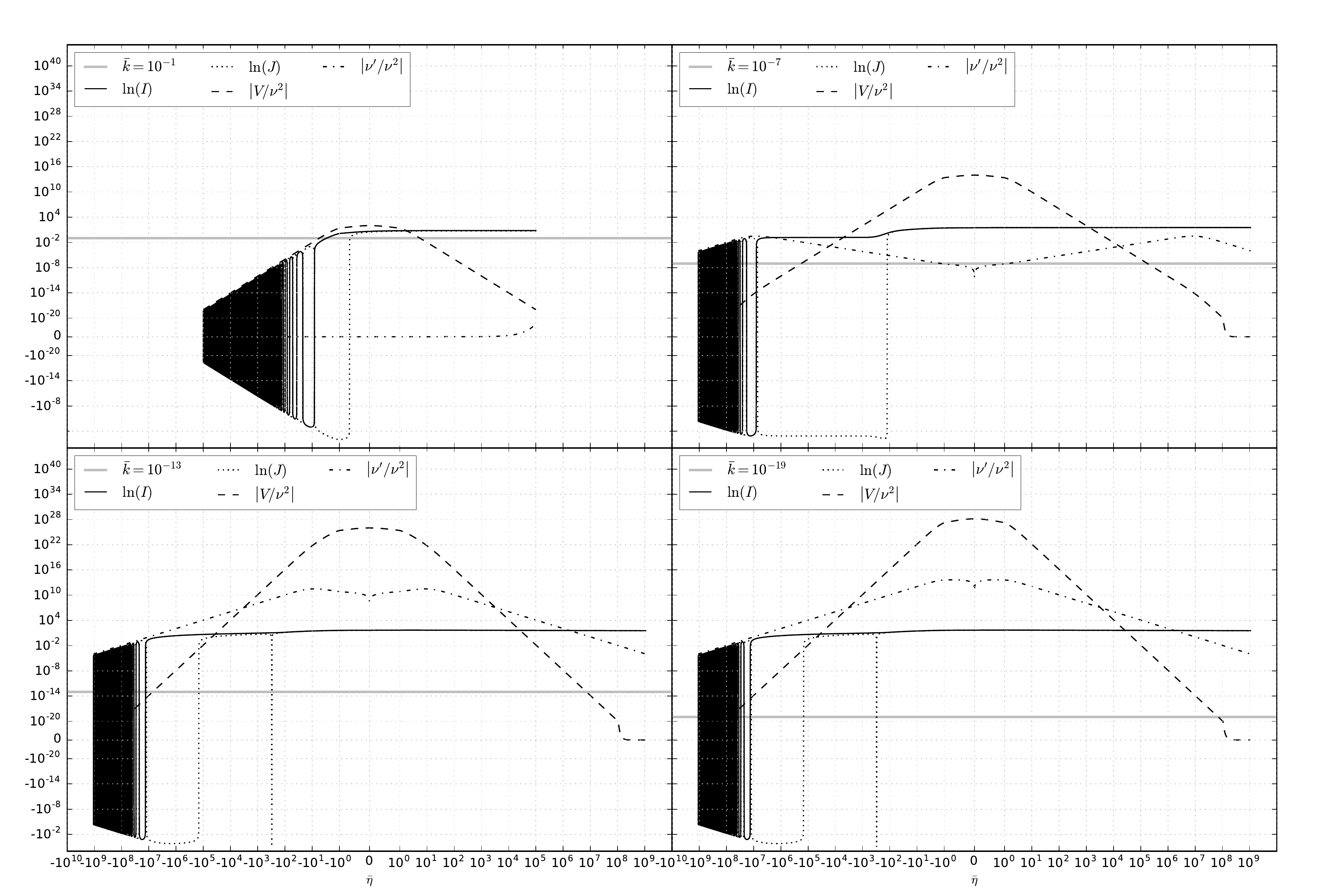}
\caption{Evolution of the adiabatic invariants for the minimal coupling,
massive case (Higgs mass), without dust, for different values of $k$. In
this situation one must consider the effect of $\nu^\prime/\nu^2$
together with $V/\nu^2$. In the top panels, the values of $\bar{k}$ are
such that $\nu^\prime/\nu^2$ is less important than $V/\nu^2$, hence, the
behavior is similar to the massless case showed in
Fig.~\ref{fig:evol:min:massless}. The lower panels, show the situation
for small $\bar{k}$. In this case, the contribution of the mass term is
important and $\nu^\prime/\nu^2$ overcome the potential term before and
after the bounce, yielding the conformal coupling spectrum for these
modes. The dominance of $V/\nu^2$ near the bounce, absent in the
conformal coupling case, makes the amplitude increases substantially with
respect to the former case. } \label{fig:evol:min:massive}
\end{figure*}

Figure~\ref{fig:RadCoeffsMinimal} shows the Bogoliubov coefficients in
the minimal coupling case with many different masses. It can be seen that
the Bogoliubov coefficients are many orders of magnitude greater than in
the conformal case, see Fig.~\ref{fig:RadCoeffsConformal}. In
Fig.~\ref{fig:evol:min:massive}, we can see examples of the power-laws
present in the Bogoliubov coefficients. For small $\bar{k}$, the spectrum
has its shape similar to the massive conformal coupling case, but the
amplitude is increased by a factor of $\pi^2/r_b$ (this factor was found
by fitting the small $\bar{k}$ interval with a power-law). This can be
understood looking at the lower panels in
Fig.~\ref{fig:evol:min:massive}. In these plots, it is evident that the
$\nu^\prime/\nu^2$ term dominates earlier, and controls the way the field
leaves the adiabatic regime. Nevertheless, around the bounce phase, and
differently from the conformal coupling case, the term $V/\nu^2$
dominates and increases substantially the amplitude of the adiabatic
invariants. For larger $\bar{k}$, the maximum of $\nu^\prime/\nu^2$ is
much smaller than that of $V/\nu^2$ [see Eqs.~\eqref{eq:etam}].
Therefore, in these cases it is the $V/\nu^2$ the one which halts the
adiabatic regime and, consequently, the spectrum ends up similar to the
massless case.

Finally, as discussed in Sec.~\ref{sec:massless}, the total particle
number density for the massive and massless are equal, within the
required precision, and they are both given by Eq.~\eqref{eq:n:min}. In
the ultra-relativistic limit, the energy density is given by
Eq.~\eqref{eq:rhoc:min}. In the non-relativistic limit, the energy
density is described by a dust like fluid [Eq.~\eqref{eq:massive}],
namely
\begin{equation}
\Omega_\phi = \frac{m n}{\rho_{\mathrm{crit}0}} = x^3\left(\frac{m}{m_H}\right)\left(\frac{x_b}{10^{30}}\right)^3\;2.1\times 10^{6}.
\end{equation}
Hence, for the Higgs particle, any bouncing model of this type with $x_b
> 10^{27}$ will produce a dust-like fluid of Higgs particles with energy
density larger than the critical density today.

\section{Bouncing with Two Fluids: Radiation and Dust}
\label{sec:twofluids}

In this section, we consider a model of the universe where its energy
content is dominated by two fluids, dust and radiation, such that the
radiation fluid dominates during the bounce and the dust fluid dominates
in the far past before the bounce, and in the far future after the
bounce. This is a more realistic bouncing model, not only because it
takes into account the observed dark matter distribution in the universe,
but also because the matter domination epoch can account for the almost
scale invariant spectrum of cosmological perturbations indicated by
recent observations \cite{PlanckCollaboration2014,
PlanckCollaboration2015b}.

The solution for the scale factor is given by Eq.~\eqref{atnp}. Using the interval for $x_b$ defined at Eq.~\eqref{eq:xb:range}, we get the following interval for $\gamma_b$,
\begin{equation}\label{eq:gamma:range}
3.7 \times 10^{-29} < \gamma_b \ll 7.5 \times 10^{-9},
\end{equation}
where we used the values for $\Omega_{m0}$ and $\Omega_{r0}$ discussed in Sec.~\ref{sec:parprod}. To understand the effect of adding a dust-like fluid to the background model, we first study the potential term $V/\nu^2$, which dictates the mode cutoff on $n^{(i,f)}_k$. This controlling term is a monotonically decreasing function of $\bar{\eta}$ [see Eq.~\eqref{V}], and has its maximum at $\bar{\eta} = 0$,
\begin{equation}
\left.\frac{V}{\nu^2}\right\vert_{\bar{\eta}=0} = \frac{1 + 2\gamma_b}{\bar{k}^2 + r_b^2}.
\end{equation}
Within the allowed interval~\eqref{eq:gamma:range}, we see that the
presence of dust does not change significantly the maximum, which will be
approximately the same for this whole interval. Therefore, the cutoff
does not change significantly in the presence of dust.

\begin{figure}
\begin{center}
\includegraphics[scale=0.47]{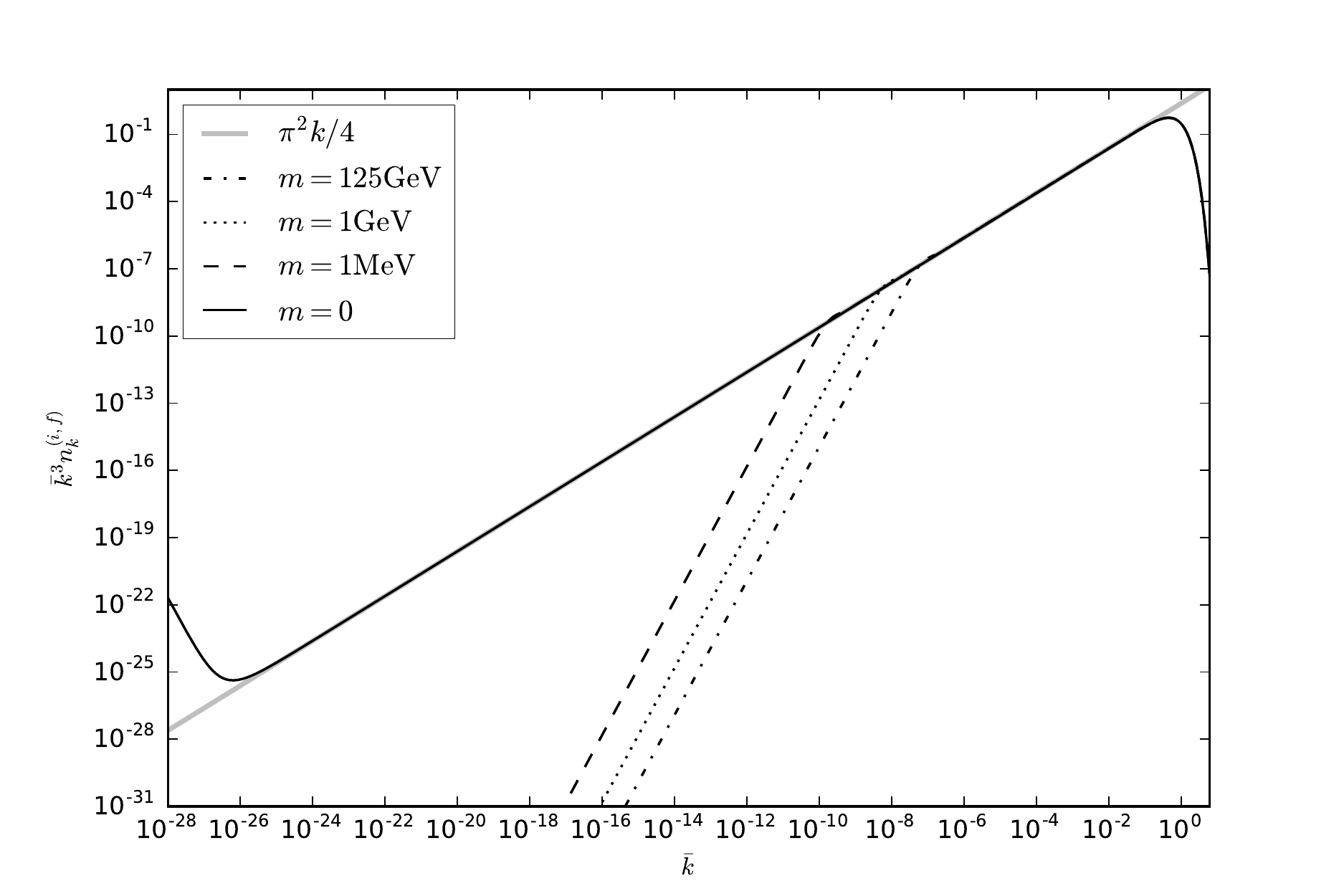} 
\end{center}
\caption{The Bogoliubov coefficients in the minimal coupling case with
the Higgs mass ($r_{b} = 8.4\times 10^{-15}$), $1\mathrm{GeV}$ mass
($r_b=6.7\times10^{-17}$), $1\mathrm{MeV}$ mass
($r_b=6.7\times10^{-20}$) and the massless case. In contrast with
Fig.~\ref{fig:RadCoeffsMinimal} here the background includes a dust like
fluid ($\gamma_b = 10^{-27}$). The result is similar to the radiation
only, with the notable exception of the massless case. Note that, for
very large wavelengths there is a growing slope that creates a infrared
divergence in the total particle number. }
\label{fig:RadCoeffsMinimalDust}
\end{figure}

\begin{figure}
\begin{center}
\includegraphics[scale=0.47]{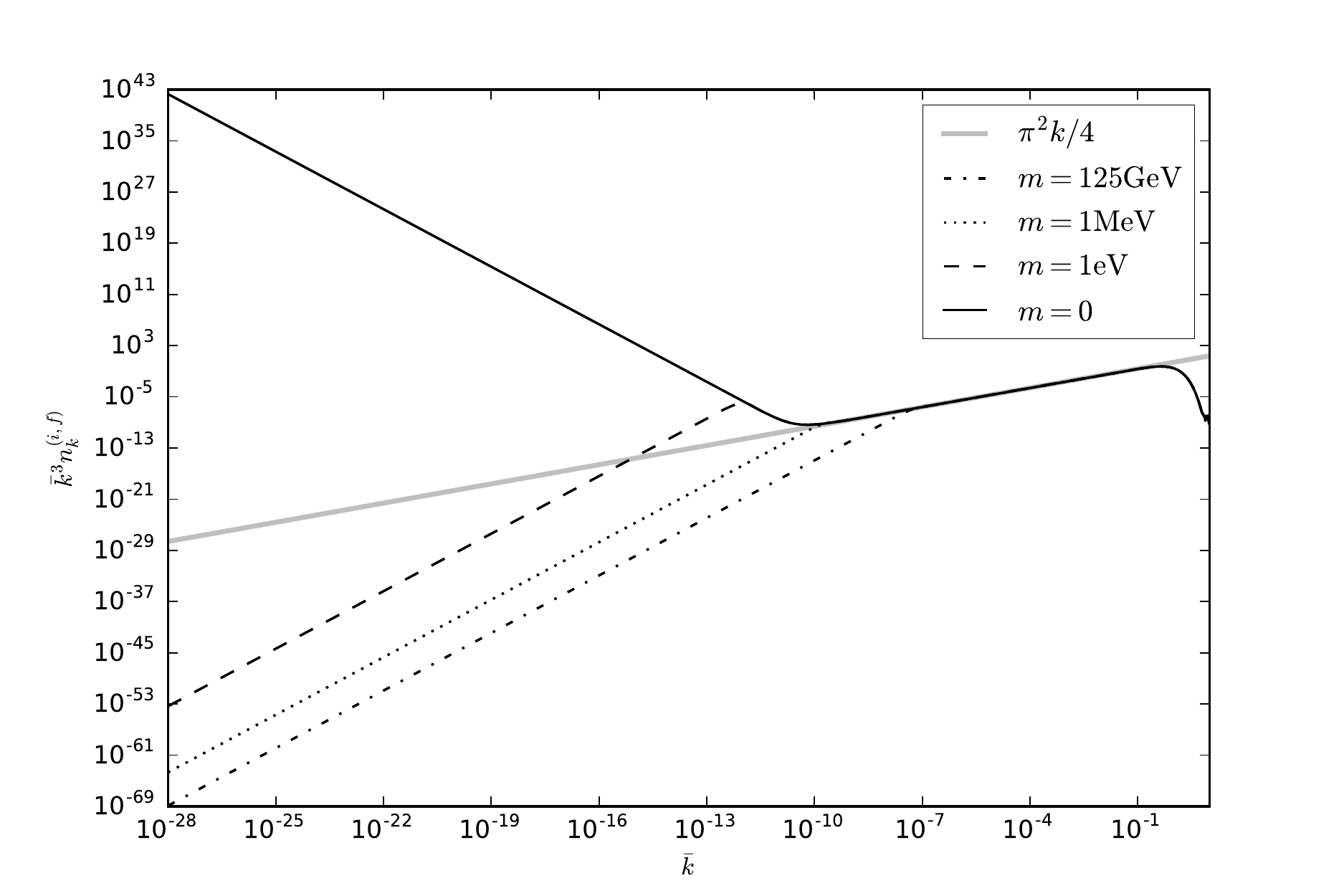} 
\end{center}
\caption{The Bogoliubov coefficients in the minimal coupling case with
the Higgs mass ($r_{b} = 8.4\times 10^{-15}$), $1\mathrm{MeV}$ mass
($r_b=6.7\times10^{-20}$), $1\mathrm{eV}$ mass ($r_b=6.7\times10^{-26}$)
and the massless case. In contrast with
Fig.~\ref{fig:RadCoeffsMinimalDust} here the background includes a dust
like fluid ($\gamma_b = 10^{-11}$) in a very shallow bounce that takes
place little earlier than the beginning of the nucleosynthesis. The
result is similar, however here the infrared divergence is much more
evident. We can also see the same behavior for the massive case, the
shape is controlled by $\nu^\prime/\nu^2$ term until the $V/\nu^2$
dominance where the form of the spectrum follows that of a massless
particle.} \label{fig:RadCoeffsMinimalHighDust}
\end{figure}

Away from the cutoff, the two controlling terms are $\nu^\prime/\nu^2$
and $V/\nu^2$, which have different intervals of importance for
different values of $\bar{k}$, $\gamma_b$ and $r_b$. For this reason,
$\beta_k^{(i,f)}$ have different forms as a function of $k$, as
appearing in Figs.~\ref{fig:RadCoeffsMinimalDust} and
\ref{fig:RadCoeffsMinimalHighDust}. For instance, in the massless case,
the infrared increase of $\beta_k^{(i,f)}$ begins for modes when
$\bar{k} < \gamma_b$, implying that the potential term be larger than
one, and the adiabatic regime terminates during this interval. We can
see this behavior in Fig.~\ref{fig:RadCoeffsMinimalDust}, where
$\bar{k}^3n_k^{(i,f)}$ changes its shape around $\bar{k} \approx
\gamma_b \approx 10^{-27}$, and in
Fig.~\ref{fig:RadCoeffsMinimalHighDust} where it changes at $\bar{k}
\approx \gamma_b \approx 10^{-11}$.

Thus, apart from the infrared divergence in the massless case, the
amount of particle creation not only does not depend on the mass of the
particles (except for the conformal case), but also does not depend on
the radiation-matter equality constant $\gamma_b$, for values within the
constraint \eqref{eq:gamma:range}. The Figs.~\ref{fig:RadCoeffsMinimal},
\ref{fig:RadCoeffsMinimalDust} and \ref{fig:RadCoeffsMinimalHighDust}
show that gravitational particle creation is effective mostly near the
bounce, which in all cases is dominated by the radiation fluid. In
particular, the fluid that dominates before radiation is not important.
All that matters is the gravitational coupling and which fluid dominates
during the bouncing phase. If, for instance, we bring the moment of
radiation-matter equality close enough to the bounce, as in
Fig.~\ref{fig:RadCoeffsMinimalHighDust}, then the amount of particle
creation becomes sensitive to $\gamma_b$. However, this happens only for
values far to the left outside the constraint \eqref{eq:gamma:range},
i.e. for $\gamma_b \gg 10^{-9}$. Hence, the results of the last section
for the number density of particles created and their energy densities
are the same.

The infrared divergence of $n_k^{(i,f)}$, in the massless case, happens
for particles with very long wave-lengths. For example, in
Fig.~\ref{fig:RadCoeffsMinimalDust}, it starts at $\bar{k} \approx
10^{-27}$, which is approximately $0.4\;\mathrm{Mpc}$, but it begins with
very small amplitude $\approx 10^{-25}$ and, consequently, it will be
only relevant for much larger wave-lengths
($\approx10^{8}\,\mathrm{Gpc}$). These modes leave the adiabatic regime
very early $\bar{k}\bar{\eta} \approx 1$, so any contracting model with a
finite initial time $\bar{\eta}_i$ have a maximum mode infrared cutoff
$\bar{k}_i = \bar{\eta}_i^{-1}$. In these cases, the adiabatic vacuum is
not well defined for long wave-lengths. Nonetheless, these modes
represent wave-lengths larger than the initial universe scale, and one
could simply choose to suppress these non-causal modes. Also, if the
universe is finite in volume, then the allowed modes $\bar{k}$ would be
discrete, and the infrared divergence would disappear, since in this case
we have a maximum allowed wave-length. Alternatively, this divergence can
be seen as a constraint on the initial conditions for the contracting
model.

Finally, it is worth pointing out that the potential term responsible for
this infrared divergence is the same used to produce an almost invariant
power-spectrum for the cosmological perturbations in the matter-bounce
(see, for example~\cite{Novello2008,Battefeld2015}). Therefore, the
matter-bounce scenario naturally leads to a heavy production of long
wave-length particles, unless constraints on their size on time and/or
space are imposed.

\section{Conclusion}
\label{conclusion}

In this paper, we have calculated the energy density resulting from
quantum particle creation in bouncing models with scale factors given by
Eq.~\eqref{eq:aR} for the case of models with a single radiation fluid,
and Eq.~\eqref{atnp} for the case of models with dust and radiation. We
have considered the conformal and minimal coupling cases, and massive and
massless scalar fields. The results can be summarized as follows: for the
conformal coupling case, where the last term of Eq.~\eqref{eq:KG} is
absent, particle production is very small and many orders of magnitude
smaller than in the minimal coupling case, where the last term is
present. In the minimal coupling case, the energy density of massive
particles produced can be as important as the background energy density,
depending on the mass of the particle and the ratio between the scale
factor today and the scale factor at the bounce, $x_b = a_0/a_b$: for
sufficiently massive particles and deep bounces, this energy density can
exceed the background energy density. In the case of massless particles,
the energy density of produced particles can become comparable to the
background energy density only for bounces occurring at energy scales
comparable to the Planck scale or above, which lies beyond the scope of
this paper: we expect that the simple Wheeler-DeWitt approach we are
using should be valid only at scales some few orders of magnitude above
the Planck scale. Finally, the amount of particle production is almost
independent whether the model contains dust or not, as long as the fluid
which dominates at the bounce is radiation in both situations
($\gamma_b\ll10^{-9}$), except for the infrared divergence in the
massless case in the two fluids model.

These results point to the same physical conclusion: apart from the
peculiar case where the particle mass is comparable with the bounce
energy scale, particle production is only effective at the bounce itself.
That is why in the conformal coupling case, where the part of the
potential characterizing the bounce itself (the last term of
Eq.~\eqref{eq:KG}) is absent, the particle production is negligible. This
also explains why the two fluids case gives the same result as the one
fluid case, as long as the same fluid dominates at the bounce (in the
above examples, radiation). Whether the era far from the bounce is
dominated by dust or radiation does not matter for particle production.
The most relevant quantities for the number of particles created are the
coupling $\xi$ and the depth of the bounce, characterized by the number
$x_b = a_0 / a_b$: deeper the bounce, higher the number of particles
created, as seen in Eq.~\eqref{eq:n:min}.

Finally, in the massless case with two fluids, there is an infrared
divergence which, however, is important only for wavelengths much larger
than the size of the universe. Hence, for realistic models which are
spatially closed, or whose homogeneous contraction begins in a finite
time, this strong infrared production should not be present. Note that
the same infrared divergence must be present in the calculation of
cosmological perturbations in such backgrounds, as they are massless.
Hence, the same cautious remarks must be made in this situation.


\section*{ACKNOWLEDGMENTS}

Two of us (D.C.F.C. and N.P.N.) would like to thank CNPq of Brazil for
financial support, and S.D.P.V would like to thank the financial aid of
the CNPq PCI/MCTI/CBPF program.

\appendix

\section{Action Angle Variables}
\label{app:AAvar}

The equation of motion \eqref{eom} describes the evolution of the modes
for the field quantization. Its solutions oscillates when the positive
terms ($\bar{k}^2$ or the mass term) dominate over the potential $V$.
During this period of the evolution, the highly oscillatory behavior
forbid a precise numerical calculation~\cite{Iserles2002a}.

The usual workaround for this problem is to use a WKB approximation up
to a point with less oscillations, and then change for a numerical
evaluation of the solutions. Although it is also possible to work with
the WKB approximation passing from the oscillatory to the
non-oscillatory regimes, this approach is cumbersome and can lead to
wrong conclusions if care is not taken~\cite{Martin2003}. Finally, the
WKB approximation describes the solution in terms of the positive and
negative solutions separately. However, we are interested in the growth
of the negative frequency solutions whenever we start with only positive
ones.

For the reasons above we use the Action Angle (AA) variables approach,
originally used in~\cite{Vitenti2015, Peter2016} in this context. Here
we argue that this methodology provides both an approximation method and
a better suited system of equations to solve numerically. It is also
particularly convenient for the computation of particle creation, since
it describes both the positive and the negative frequencies solution
within the same approximation scheme.

The AA variables are related to the modes through the expressions
\begin{equation}
q_k = \sqrt{\frac{2I}{\nu}}\sin\theta,\qquad q_k^\prime =\sqrt{2I\nu}\cos\theta,
\end{equation}
where $q_k$ are real solutions of~\eqref{eom}. Using these variables, the
equations of motion are recast as
\begin{align}\label{eq:theta}
\theta^\prime &= \nu - \frac{V}{\nu} \sin^2\theta + \frac{\nu^\prime}{2\nu}\sin2\theta,\\ \label{eq:I}
(\ln{}I)^\prime &= -\frac{\nu^\prime}{\nu}\cos{2\theta} + \frac{V}{\nu}\sin2\theta.
\end{align}
Note that this choice of variables decouples the equations, i.e. the evolution of the angle variable $\theta$ is independent from the adiabatic invariant $I$. Thus, the integral solution for $I$ is simply,
\begin{equation}\label{eq:Isol}
I = I_0 \exp\left[{-\int_{\eta_0}^{\eta}\!\dd\eta_1\left(\frac{\nu_1^\prime}{\nu_1}\cos{2\theta_1} - \frac{V_1}{\nu_1}\sin2\theta_1\right)}\right].
\end{equation}
To build a complex solution, we introduce another set of AA variables $J$ and $\psi$, satisfying the same set above, i.e.,
\begin{align}\label{eq:psi}
\psi^\prime &= \nu - \frac{V}{\nu} \sin^2\psi + \frac{\nu^\prime}{2\nu}\sin2\psi,\\ \label{eq:J}
(\ln{}J)^\prime &= -\frac{\nu^\prime}{\nu}\cos{2\psi} + \frac{V}{\nu}\sin2\psi,
\end{align}
from which we introduce another real field variable
$$ v_k = \sqrt{\frac{2J}{\nu}}\sin\psi,\qquad v_k^\prime =\sqrt{2J\nu}\cos\psi.$$
Thus, the complex solution can be written as
\begin{align}
\phi_k = \frac{q_k+iv_k}{2i} = \frac{1}{i\sqrt{2\nu}}\left(\sqrt{I}\sin\theta + i\sqrt{J}\sin\psi\right), \\
\phi_k^\prime = \frac{q_k^\prime+iv_k^\prime}{2i} = \frac{1}{i}\sqrt{\frac{\nu}{2}}\left(\sqrt{I}\cos\theta + i\sqrt{J}\cos\psi\right).
\end{align}
The mode normalization conditions 
$$
i\left(\phi_k^*\phi_k^\prime - \phi_k^{*\prime}\phi_k\right) = 1,
$$ 
imply
\begin{equation}\label{normcond}
\sqrt{IJ}\sin\left(\psi-\theta\right) = 1.
\end{equation}
Note that the normalization condition is proportional to the Wronskian of the real and imaginary solutions, i.e., 
$$ 
\Wrons{q_k}{v_k} = 2\sqrt{IJ}\sin\left(\psi-\theta\right) = 2.
$$

Considering the adiabatic limit ($V\to0 \;\;\text{and}\;\; \nu^\prime/\nu\to0$), the solutions are simply
\begin{equation}
\begin{split}
I &= I_0, \quad J = J_0, \\ 
\theta &= \theta_0 + \int\!\dd\eta\,\nu,\quad \psi = \psi_0 + \int\!\dd\eta\,\nu.
\end{split}
\end{equation}
In this limit, the choices $\psi_0 = \theta_0 + \pi/2$ and $I_0 = J_0 =
1$ satisfy the normalization condition in Eq.~\eqref{normcond}, and we obtain the correct adiabatic vacuum, namely
\begin{equation}\label{eq:asymp}
\phi_k =  \frac{1}{\sqrt{2\nu}}\exp\left(-i\int\!\dd\eta\,\nu\right).
\end{equation}
It is worth emphasizing that in the ultraviolet limit ($k \to \infty$), the same approximate solution above applies.

The potential $V$ goes to zero in the limits $\eta \to \pm\infty$,
therefore, we can choose two initial conditions, each one matching the
adiabatic vacuum in the limit. Looking at the integral solution in
Eq.~\eqref{eq:Isol}, it is easy to see that the solutions
\begin{align}
I^{(i)} &= \exp\left[{-\int_{-\infty}^{\eta}\!\dd\eta_1\left(\frac{\nu_1^\prime}{\nu_1}\cos{2\theta_1^{(i)}} - \frac{V_1}{\nu_1}\sin2\theta_1^{(i)}\right)}\right], \\
J^{(i)} &= \exp\left[{-\int_{-\infty}^{\eta}\!\dd\eta_1\left(\frac{\nu_1^\prime}{\nu_1}\cos{2\psi_1^{(i)}} - \frac{V_1}{\nu_1}\sin2\psi_1^{(i)}\right)}\right],
\end{align}
and the condition 
$$
\lim_{\eta\to-\infty}\psi^{(i)} - \theta^{(i)} = \pi/2,
$$ 
produce the right solution matching the asymptotic in the $\eta\to-\infty$ limit, namely, $\phi^{(i)}_k$. Analogously, for the $\eta\to+\infty$ limit we obtain the solution $\phi^{(f)}_k$ through the following AA variables
\begin{align}
I^{(f)} &= \exp\left[{\int^{\infty}_{\eta}\!\dd\eta_1\left(\frac{\nu_1^\prime}{\nu_1}\cos{2\theta_1^{(f)}} - \frac{V_1}{\nu_1}\sin2\theta_1^{(f)}\right)}\right], \\
J^{(f)} &= \exp\left[{\int^{\infty}_{\eta}\!\dd\eta_1\left(\frac{\nu_1^\prime}{\nu_1}\cos{2\psi_1^{(f)}} - \frac{V_1}{\nu_1}\sin2\psi_1^{(f)}\right)}\right],
\end{align}
and the condition 
$$
\lim_{\eta\to\infty}\psi^{(f)} - \theta^{(f)} = \pi/2.
$$

Both solutions $\phi_k^{(i)}$ and $\phi_k^{(f)}$ provide a well defined adiabatic vacuum for $\vert\eta\vert \gg 1$ accordingly. In these intervals $V / \nu^2 \ll 1$ and $\nu^\prime/\nu \ll 1$ and consequently $$ I \approx 1,\qquad J \approx 1,\qquad \psi - \theta \approx \pi/2.$$ Particularly, this means that both solutions remain close to the form given in Eq.~\eqref{eq:asymp} during this time interval.

Finally, we need to compute the products between these two solutions. The Bogoliubov coefficients are given by,
\begin{align}
\alpha_k^{(i,f)} &= i\left(\phi_k^{(i)*}\phi_k^{(f)\prime} - \phi_k^{(i)\prime*}\phi_k^{(f)}\right), \\
\beta_k^{(i,f)} &= i\left(\phi_k^{(i)}\phi_k^{(f)\prime} - \phi_k^{(i)\prime}\phi_k^{(f)}\right).
\end{align}
Writing these expression in terms of the real solutions we get,
\begin{equation}\label{eq:alpha}
\begin{split}
\alpha_k^{(i,f)} = &-\frac{1}{2}\left[\Wrons{q_k^{(i)}}{v_k^{(f)}} + \Wrons{q_k^{(f)}}{v_k^{(i)}}\right]\\ 
&+\frac{i}{2}\left[\Wrons{q_k^{(i)}}{q_k^{(f)}} + \Wrons{v_k^{(f)}}{v_k^{(i)}}\right],
\end{split}
\end{equation}
and
\begin{equation}\label{eq:beta}
\begin{split}
\beta_k^{(i,f)} = &+\frac{1}{2}\left[\Wrons{q_k^{(i)}}{v_k^{(f)}} - \Wrons{q_k^{(f)}}{v_k^{(i)}}\right]\\
&-\frac{i}{2}\left[\Wrons{q_k^{(i)}}{q_k^{(f)}} - \Wrons{v_k^{(f)}}{v_k^{(i)}}\right].
\end{split}
\end{equation}
Since these Wronskians are constant we can evaluate them at any time. In the limit $\eta_f \gg 1$ we obtain,
\begin{align}\label{eq:alpha2}
\left\vert\alpha_k^{(i,f)}\right\vert^2 = \left.\frac{1}{4}\left(I^{(i)}+J^{(i)} + 2\right)\right\vert_{\eta=\eta_f},\\ \label{eq:beta2}
\left\vert\beta_k^{(i,f)}\right\vert^2 = \left.\frac{1}{4}\left(I^{(i)}+J^{(i)}-2\right)\right\vert_{\eta=\eta_f}.
\end{align}
The result above shows the convenience of this approach. The adiabatic
invariants $I^{(i)}$ and $J^{(i)}$ provide the value of $\beta_k^{(i,j)}$
when evaluated at $\eta_f$. It is also worth noting that although
$I^{(i)}$ and  $J^{(i)}$ are adiabatic invariants, their logarithm is
not. In the initial regime $\bar{\eta} \ll -1$, $I^{(i)}$ can be
approximated as
\begin{equation}
I^{(i)} \approx 1 -\int_{-\infty}^{\eta}\!\dd\eta_1\left(\frac{\nu_1^\prime}{\nu_1}\cos{2\theta_1^{(i)}} - \frac{V_1}{\nu_1}\sin2\theta_1^{(i)}\right),
\end{equation}
showing that, as long as $V/\nu \ll 1$ and $\nu^\prime/\nu \ll 1$, $I^{(i)} \approx 1$ provides a good approximation for this variable, where the same reasoning applies to $J^{(i)}$.

During the oscillatory regime, we can also obtain the first order approximation of the integrals $I^{(i)}$ and $J^{(i)}$ as~\cite{Erdelyi1956},
\begin{align}\label{eq:approx:I}
\ln\left(I^{(i)}\right) &\approx -\frac{\nu^\prime}{2\nu^2}\sin{2\theta^{(i)}} - \frac{V}{2\nu^2}\cos2\theta^{(i)}, \\ \label{eq:approx:J}
\ln\left(J^{(i)}\right) &\approx -\frac{\nu^\prime}{2\nu^2}\sin{2\psi^{(i)}} - \frac{V}{2\nu^2}\cos2\psi^{(i)}.
\end{align}
Similarly, for $\theta^{(i)}$ and $\psi^{(i)}$ we get
\begin{align}
\theta^{(i)} &\approx \sigma - \frac{\nu^\prime}{4\nu^2}\cos{2\theta^{(i)}} + \frac{V}{4\nu^2}\sin2\theta^{(i)}, \\ \label{eq:approx:psi}
\psi^{(i)} &\approx \frac{\pi}{2} + \sigma - \frac{\nu^\prime}{4\nu^2}\cos{2\psi^{(i)}} + \frac{V}{4\nu^2}\sin2\psi^{(i)},
\end{align}
where
\begin{equation}
\sigma = \int\dd\eta\,\nu\left(1-\frac{V}{\nu^2}\right).
\end{equation}

In the numerical calculations, we used the approximations above. We set
the initial time for each mode at the point $\bar{\eta} < -1$ where
\begin{equation}\label{eq:condition}
\mathrm{MAX}\,(\vert{}V/\nu^2\vert,\vert\nu^\prime/\nu^2\vert) = \epsilon,
\end{equation}
where $\epsilon$ controls the precision required.

During the non-adiabatic evolution, we can compute the solutions for Eqs.~\eqref{eq:theta} and \eqref{eq:I} in two cases. Ignoring all terms but the $\nu^\prime/\nu$ we get:
\begin{align}
\theta &= \cot^{-1}\left(\cot\theta_0 \frac{\nu_0}{\nu}\right), \\ \label{eq:evol:I:nu}
I &= I_0\left(\sin^2\theta_0\frac{\nu}{\nu_0} + \cos^2\theta_0\frac{\nu_0}{\nu}\right).
\end{align}
When evolving through the bounce phase, since $\nu$ is an even function
and considering $\eta_0 < 0$, we obtain $I(-\eta_0) = I(\eta_0)$, where
subindex ${}_{0}$ denotes some arbitrary initial time. In short, for an
even $\nu$, the evolution through the bounce for modes for which
$\nu^\prime/\nu^2$ dominates, makes $I$ return to the same value it
started.

The second case, where the term $V/\nu$ is the only relevant one, have
the following solutions
\begin{align}\label{eq:noaiab:theta}
\theta &= \cot^{-1}\left[\cot\theta_0 + f(\eta,\eta_0)\right], \\ \label{eq:noaiab:I}
I &= I_0\left[1 + f(\eta,\eta_0)\left(f(\eta,\eta_0)\sin^2\theta_0+\sin2\theta_0\right)\right],
\end{align}
where 
$$
f(\eta,\eta_0) = \int_{\eta_0}^{\eta}\!\dd\eta_1\frac{V_1}{\nu_1}.
$$
The solution above is valid for both $I$ and $J$. Remembering that
$\theta$ and $\psi$ have a $\pi/2$ difference, if $\theta_0 = 0$, then
$I = I_0$ and $J = J_0[1 + f(\eta,\eta_0)^2]$. Therefore, the adiabatic
invariants have their amplitude increased by a factor of
$1+f(\eta,\eta_0)^2$.

\section{WKB Method}
\label{app:WKBM}

For the sake of completeness, we have also done numerical calculations
using WKB approximations. This has a two-fold purpose: first, to compare
the results with those obtained by the method of AA variables; second, to
give a clearer picture of the process, since the WKB method is both
well-known and intuitive, when applicable. In this approach the first
order WKB approximation is used to provide the initial conditions for the
mode equations near the end of the oscillatory phase. This is crucial to
obtain a well posed numerical evolution, should we start too early in the
high oscillatory phase the numerical evolution would be slow and possibly
dominated by the global cumulative error~\cite{Iserles2002a}. Moreover,
evolving the solution in the expanding phase, after it reenters the
oscillatory regime, would cause the same problem. For this reason, at
this point we define the WKB approximation of the future vacuum.  Then,
using this mode function, the Bogoliubov coefficient can be calculated at
the point early after the numerical solution reenters the oscillatory
phase.

Here, we consider only the minimal coupling cases. Since the $\xi$
dependent term in the Klein-Gordon equation \eqref{eq:KG} vanishes for
any value of $\xi$ in the asymptotic past and future, we use for initial
conditions the asymptotic solutions with conformal coupling, which will
also render a vacuum definition in the past. Since the scale factor is
symmetric in both cases, one and two-fluid models, an out-vacuum can then
be defined in the future to be compared to the in-vacuum.

In the one-fluid model, with scale factor \eqref{eq:aR}, the asymptotic
solutions of Eq.~\eqref{eq:KG:r} are Eqs.~\eqref{eq:exact:sol} and
\eqref{eq:exact:cc}, for massive fields. In this case, we also used the
first order WKB instead of the analytical asymptotic solution, obtaining
the same results. For massless particles, the first order WKB solutions
are simple plane-waves~\eqref{eq:KG:pw}. We use, for initial time,
$\eta_i < 0$ such that
\begin{equation}\label{eq:intime}
\epsilon = \frac{1}{\left[\bar{k}^{2} + r_{b}^{2} \left( 1 + \bar{\eta}_{i}^{2} \right) \right]\left(1 + \bar{\eta}_{i}^{2} \right)^{2}},
\end{equation}
where we chose $\epsilon = 10^{-6}$, and we evolved the solutions until
$\bar{\eta}_{f} = - \bar{\eta}_{i}$. This choice of initial/final time
controls how deep in the oscillatory phase we begin/end the numerical
calculation. The initial conditions are given by \eqref{eq:exact:sol} and
\eqref{eq:exact:cc} for massive fields, or \eqref{eq:KG:pw} for massless,
at $\bar{\eta}_i$. Then, we use equation \eqref{bogocoef} to calculate
the Bogoliubov coefficients.

In the two-fluid model, we follow an analogous procedure. Now, however,
we obtained for the first order WKB solutions to equation \eqref{eq:KG}
with conformal coupling and scale factor \eqref{atnp}, for massive
particles,
\begin{widetext}
\begin{align} \label{eq:raddust:wkb:in}
 \f^{(i)}_{k} (\bar{\eta}) & = \left( 2 \sqrt{\bar{k}^{2} + r_{b}^{2} \gamma_{b}^{2} \bar{\eta}^{4}} \right)^{-1/2} \mbox{exp} \left\{ -\frac{i \bar{\eta}}{3} \sqrt{\bar{k}^{2} + r_{b}^{2} \gamma_{b}^{2} \bar{\eta}^{4}} \left[1 + 2 \hs{1mm} {}_{2}F_{1}\left(\frac{3}{4},1;\frac{5}{4}; -\frac{r_{b}^{2} \gamma_{b}^{2} \bar{\eta}^{4}}{\bar{k}^{2}} \right) \right] \right\},
 \hspace{0.5cm} \bar{\eta} < 0, \\
 \label{eq:raddust:wkb:out}
 \f^{(f)}_{k} (\bar{\eta}) & = \left( 2 \sqrt{\bar{k}^{2} + r_{b}^{2} \gamma_{b}^{2} \bar{\eta}^{4}} \right)^{-1/2} \mbox{exp} \left\{ -\frac{i \bar{\eta}}{3} \sqrt{\bar{k}^{2} + r_{b}^{2} \gamma_{b}^{2} \bar{\eta}^{4}} \left[1 + 2 \hs{1mm} {}_{2}F_{1}\left(\frac{3}{4},1;\frac{5}{4}; -\frac{r_{b}^{2} \gamma_{b}^{2} \bar{\eta}^{4}}{\bar{k}^{2}} \right) \right] \right\}, \hspace{0.5cm} \bar{\eta} > 0,
\end{align}
\end{widetext}
where ${}_{2}F_{1}$ is an Hypergeometric function~\cite{Gradshteyn2014}.
While, for massless particles, we can still use \eqref{eq:KG:pw}.

Using the WKB method, we found the same results as with the action
angle variables approach. Here we used the WKB mode functions to define
the initial conditions and as approximations of the future vacuum mode
functions. For the time interval where the WKB approximation is not
valid, Eq.~\eqref{eom} was used to calculate numerically the time
evolution. Finally, we compare the numerical solution with the future
defined WKB approximation in order to obtain the Bogoliubov coefficients.
In contrast, in the AA approach we used the first order adiabatic
approximations given in Eqs.~(\ref{eq:approx:I}--\ref{eq:approx:psi}) as
initial condition in the oscillatory phase, and then, we evolved
Eqs.~\eqref{eq:theta}, \eqref{eq:I}, \eqref{eq:psi} and \eqref{eq:J}
numerically until $I$ and $J$ freeze out to their final values.

\bibliography{references}
\bibliographystyle{apsrev4-1}

\end{document}